\documentclass[preprint]{aastex63}
\usepackage{graphics}
\usepackage{mathrsfs}
\usepackage{comment}
\usepackage{amsmath}
\usepackage{bashful}
\usepackage{float}
\usepackage{threeparttable}
\newcommand{\com}{\color{red}}

\submitjournal{ApJ}
\received{August 15}
\revised{November 12}
\accepted{November 15, 2022}

\shorttitle{Photosynthetic fluorescence on Exoplanets}
\shortauthors{Komatsu et al.}

\begin{document}

\title{Photosynthetic Fluorescence from Earth-Like Planets around Sun-Like and Cool Stars}

\correspondingauthor{Yu Komatsu}
\email{yu.komatsu@nao.ac.jp}

\author[0000-0002-0371-2885]{Yu Komatsu}
\affiliation{Astrobiology Center, 2-21-1 Osawa, Mitaka, Tokyo 181-8588, Japan.}
\affiliation{National Astronomical Observatory of Japan, 2-21-1 Osawa, Mitaka, Tokyo 181-8588, Japan.}

\author[0000-0003-4676-0251]{Yasunori Hori}
\affiliation{Astrobiology Center, 2-21-1 Osawa, Mitaka, Tokyo 181-8588, Japan.}
\affiliation{National Astronomical Observatory of Japan, 2-21-1 Osawa, Mitaka, Tokyo 181-8588, Japan.}

\author[0000-0002-4677-9182]{Masayuki Kuzuhara}
\affiliation{Astrobiology Center, 2-21-1 Osawa, Mitaka, Tokyo 181-8588, Japan.}
\affiliation{National Astronomical Observatory of Japan, 2-21-1 Osawa, Mitaka, Tokyo 181-8588, Japan.}

\author[0000-0002-7415-1536]{Makiko Kosugi}
\affiliation{Astrobiology Center, 2-21-1 Osawa, Mitaka, Tokyo 181-8588, Japan.}
\affiliation{National Astronomical Observatory of Japan, 2-21-1 Osawa, Mitaka, Tokyo 181-8588, Japan.}
\affiliation{National Institute for Basic Biology, 38 Nishigonaka, Myodaiji, Okazaki, Aichi 444-8585, Japan.}

\author[0000-0002-1060-5738]{Kenji Takizawa}
\affiliation{Astrobiology Center, 2-21-1 Osawa, Mitaka, Tokyo 181-8588, Japan.}
\affiliation{National Institute for Basic Biology, 38 Nishigonaka, Myodaiji, Okazaki, Aichi 444-8585, Japan.}

\author[0000-0001-8511-2981]{Norio Narita}
\affiliation{Komaba Institute for Science, The University of Tokyo, 3-8-1 Komaba, Meguro, Tokyo 153-8902, Japan.}
\affiliation{Astrobiology Center, 2-21-1 Osawa, Mitaka, Tokyo 181-8588, Japan.}
\affiliation{Instituto de Astrof\'{i}sica de Canarias (IAC), 38205 La Laguna, Tenerife, Spain}

\author[0000-0002-5051-6027]{Masashi Omiya}
\affiliation{Astrobiology Center, 2-21-1 Osawa, Mitaka, Tokyo 181-8588, Japan.}
\affiliation{National Astronomical Observatory of Japan, 2-21-1 Osawa, Mitaka, Tokyo 181-8588, Japan.}

\author[0000-0003-3380-3953]{Eunchul Kim}
\affiliation{National Institute for Basic Biology, 38 Nishigonaka, Myodaiji, Okazaki, Aichi 444-8585, Japan.}

\author[0000-0001-9194-1268]{Nobuhiko Kusakabe}
\affiliation{Astrobiology Center, 2-21-1 Osawa, Mitaka, Tokyo 181-8588, Japan.}
\affiliation{National Astronomical Observatory of Japan, 2-21-1 Osawa, Mitaka, Tokyo 181-8588, Japan.}

\author[0000-0002-1386-1710]{Victoria Meadows}
\affiliation{Department of Astronomy and Astrobiology Program, University of Washington, Box 351580, Seattle, Washington 98195, USA.}
\affiliation{NASA Nexus for Exoplanet System Science, Virtual Planetary Laboratory Team, Box 351580, University of Washington, Seattle, Washington 98195, USA.}

\author[0000-0002-6510-0681]{Motohide Tamura}
\affiliation{Astrobiology Center, 2-21-1 Osawa, Mitaka, Tokyo 181-8588, Japan.}
\affiliation{National Astronomical Observatory of Japan, 2-21-1 Osawa, Mitaka, Tokyo 181-8588, Japan.}
\affiliation{Department of Astronomy, Graduate School of Science, The University of Tokyo, 7-3-1, Hongo, Bunkyo-ku, Tokyo 113-0033, Japan.}

%\linenumbers
\begin{abstract}
Remote sensing of the Earth has demonstrated that photosynthesis is traceable as the vegetation red edge (VRE), which is the steep rise in the reflection spectrum of vegetation, and as solar-induced fluorescence.
This study examined the detectability of biological fluorescence from two types of photosynthetic pigments, chlorophylls (Chls) and bacteriochlorophylls (BChls), on Earth-like planets with oxygen-rich/poor and anoxic atmospheres around the Sun and M dwarfs.
Atmospheric absorption, such as H$_2$O, CH$_4$, O$_2$, and O$_3$, and the VRE obscure the fluorescence emissions from Chls and BChls.
We found that BChl-based fluorescence for wavelengths of 1000--1100\,nm, assuming the spectrum of BChl\,$b$-bearing purple bacteria, could provide a suitable biosignature but only in the absence of the water cloud coverage or other strong absorbers near 1000\,nm.
The Chl fluorescence is weaker for several reasons, e.g., spectral blending with the VRE.
The apparent reflectance excess is greatly increased in both Chl and BChl cases around TRAPPIST-1 due to fluorescence and stellar absorption lines.
This could be a promising feature for detecting the fluorescence around ultracool red dwarfs by follow-up ground-based observations with high spectral resolution; however, it requires a long time around Sun-like stars, even for a LUVOIR-like space mission. 
%{\com Moreover, the simultaneous detection of fluorescence and VRE, as well as the atmospheric features, is key to identifying traces of photosynthesis and avoiding false positive or negative detection because absorption, reflectance, and fluorescence are physically connected.}
Moreover, the simultaneous detection of fluorescence and VRE is key to identifying traces of photosynthesis because absorption, reflectance, and fluorescence are physically connected.
For further validation of fluorescence detection, the nonlinear response of biological fluorescence as a function of light intensity could be considered.

\end{abstract}

\section{Introduction} \label{sec:intro}

The ultimate goal of characterizing rocky planets is to identify potential biosignatures, spectral fingerprints of atmospheric gases, and surface features produced by biological activities \citep{des2002remote,schwieterman2018exoplanet, meadows2018exoplanet}.
The simultaneous identification of oxygen, ozone, and methane on rocky habitable planets shows promise as a way to detect  Earth-like life.
Oxygenic photosynthesis produces a unique feature in the reflection spectrum on a planetary surface, called the vegetation red edge (VRE), as well as biosignature gases~\citep{kiang2007spectral-b}. The VRE is the steep difference in the reflection spectrum of the surface vegetation around 700 nm due to chlorophyll (Chl) absorption in the visible region and the large reflectance by cell structures in the near-infrared (NIR) region~\citep{gates1965spectral,jacquemoud1990prospect}. 
Remote sensing of the Earth and Earthshine observations provide spectral indices involved in the VRE, such as the NDVI, which is a normalized difference in the reflection spectrum of the Earth between the visible and NIR wavelength regions. 
The Moderate Resolution Imaging Spectroradiometer (MODIS) onboard NASA's Terra satellite at 16-day intervals at 500 m and 1 km resolutions shows that the NDVI varies from ~0.05 to nearly 0.9, whose upper limit is obtained at a dense forest site during the peak growing season~\citep{huete2002overview}.
Whereas remote sensing observes local areas on Earth, Earthshine observations provide disk-averaged spectra of the Earth, leading to fruitful insights into exoplanet applications. The apparent reflectance change in the Earth's disk-averaged spectrum due to surface vegetation is  less than 2\%~\citep{montanes2006vegetation}. 
The NDVI calculated from the Earthshine observations varies up to $\sim$0.10, depending on different views of the Earth, and is reduced by cloud coverage~\citep{tinetti2006detectability}.
The application of NDVI to disk-averaged spectra assuming Earth-like exoplanets requires caution because remote sensing observes only local areas on the Earth to map vegetation. 
For instance, \cite{Livengood2011-ax} found that additional spectral bands to NDVI are required to distinguish between the Earth vegetation and the Moon surface.

The VRE signals from exoplanets around stars other than a Sun-like star are challenging to predict due to the complexity of photosynthetic mechanisms in different light environments.
However, the VRE on exoplanets may still be recognizable as an anomalous time-varying due to seasonal variability of the vegetation, and step-function-like spectroscopic feature at wavelengths different from those on the Earth~\citep{seager2005vegetation}.
\cite{tinetti2006detectability} proposed that if a three-photon photosynthetic scheme were working on exoplanets around M dwarfs, where there was little or no visible light, then the red edge of vegetation could also be shifted into the NIR.
%However, according to \cite{takizawa2017red}, the Earth-type oxygenic photosynthesis that uses visible light could be preferred to NIR using photosynthesis underwater on any exoplanet. 
However, according to \cite{takizawa2017red}, even around M dwarfs, the evolution of photosynthesis in water may drive a preference for using visible light rather than NIR, even after organisms colonize land surfaces.
Moreover, the light absorption properties of land vegetation could be optimized after long-term adaptive evolution depending on stellar irradiations as estimated by \cite{Lehmer_2021}.
%}
Anoxygenic photosynthesis as performed by organisms such as purple bacteria, is thought to precede the emergence of oxygenic photosynthesis, whose global effect was characterized by the great oxidation event ($\sim$2.3 billion years ago (Ga)). \cite{sanroma2013characterizing} discussed the detectability of light reflected from purple bacteria with bacteriochlorophyll (BChl) as a photosynthetic pigment. They showed that purple bacteria exhibit detectable features, and their VRE peak is redder than higher plants, assuming an Earth-like planet before the rise of oxygen.
In a comprehensive study of different pigment reflectivity, \cite{schwieterman2015nonphotosynthetic} showed that both nonphotosynthetic pigments and photosynthetic pigments affect the disk-averaged spectra.
Furthermore, as for false positive detection, the reflectance features of some minerals on the Earth are similar to the VRE ones~\citep{seager2005vegetation,Schwieterman2018}.
Thus, extracting the VRE signal from reflected light should require knowledge of the surface environment on an exoplanet and high-resolution spectroscopic observations.

Fluorescence is another photosynthesis-related phenomenon that could also be a remote-sensing biosignature. 
Fluorescence is one of the de-excitation processes of photosynthetic pigments from the excited states to the ground state, along with intersystem crossing and inner conversion. 
Photosynthetic organisms on the Earth use Chls or BChls as light-absorbing pigments and electron donors/acceptors in the primary reactions of photosynthesis.
The photon energy captured by Chls/BChls is mainly transferred to the reaction center (RC), which is the pigment-protein complex at the center of the photosystem used for photochemical reactions. A part of photon energy is, however, dissipated as heat or emitted as fluorescence from light-harvesting antenna systems, which are pigment-protein complexes surrounding RC that capture light energy and deliver the energy to the RC.
Excess photon energy is preferentially removed as heat dissipation, rather than fluorescence.
As a result, fluorescence yield tends to be a smaller percentage of the excess energy and fluctuates with the degree of the excitation energy transfer (EET) between Chls, and heat dissipation.
%The excitation energy transfer (EET) between Chls and heat dissipation preferentially occurs compared to fluorescence emission in photosystems. As a result, fluorescence yield fluctuates with the degree of EET and heat dissipation. 
The fluorescence yield of photosynthetic organisms is estimated to be $\sim$5\%, whereas that of free Chls/BChls in organic solvents is $\sim30$\%~\citep{grimm2006chlorophylls}. 

Plants and other oxygenic phototrophs use two different photosystems in sequence, that is, photosystem II (PSII) and photosystem I (PSI).
The energy level of the RC of PSII is higher, being equivalent to 680\,nm, than that of PSI.
In general, Chl fluorescence is mainly emitted from PSII because the excess light energy in PSI is immediately dissipated as heat. 
Therefore, the fluorescence spectrum of a cell has a peak at 680\,nm, and the distribution of fluorescence emission extends to wavelengths up to 780\,nm.
Note that fluorescence emissions at 680\,nm under highly concentrated Chls conditions, such as a leaf structure, decrease due to reabsorption by peripheral Chls with a red-absorption band.
Conversely, the six BChls (BChl\,$a$, $b$, $c$, $d$, $e$, and $g$) used in non-oxygenic photosynthetic bacteria, such as purple bacteria, green sulfur and nonsulfur bacteria and heliobacteria~\citep{kiang2007spectral}, mainly absorb far-red light in vivo. The BChl\,$b$ in purple bacteria has the longest wavelength absorbance (1010 nm) and fluorescence (1050 nm) emissions. However, the detailed characteristics of fluorescence from BChls, such as fluorescence yield and its variation in light environments, remain poorly understood.

In contrast to the VRE which tracks the vegetation mass in the remote sensing of the Earth, fluorescence can be used as an indicator of active photosynthesis. The fluorescence signal emitted from the global ground vegetation, which is called solar-induced fluorescence (SIF), can be detected by remote sensing from satellites as excess light seen in the absorption of Fraunhofer lines in sunlight reflected from the Earth, which is the apparent increase in the reflectance spectrum due to fluorescence \citep{maier2004sun}. 
The observation of SIF is fundamentally challenging because the small SIF signal is overwhelmed by large background signals in the reflected sunlight.
Then, high-resolution spectroscopy utilizes specific wavelengths with large solar absorption, which means the low intensity of reflected light.
The SIF is observed as the in-filling effect at these wavelengths.
This methodology works because a large contrast is ensured between the Sun and the reflected light from the Earth at specific wavelengths.
Thus, SIF has been observed in absorption bands by the Fourier high-dispersion spectrometers onboard many environmental satellites (e.g., GOSAT~\citep{hamazaki2005fourier,lee2013forest}, GOME-2~\citep{callies2000gome}, and GOSAT-2~\citep{nakajima2012current}), which produce the time-series SIF map of Earth~\citep{frankenberg2014prospects,sun2018overview}. 
We can extract information on the ground vegetation and atmospheric/surface environment, especially the gross primary production (GPP), from the changes in the fluorescence map by calibrating the remote observations with the results of local ground observations~\citep{sun2018overview}.
Such as the SIF in Earth observations, the detection of photosynthetic fluorescence in a planet around stars will investigate the surface environment and vegetation conditions on exoplanets.
High-resolution spectroscopy would be inevitable for the exofluorescence detection, and the contrast between a planet and its host star should be high enough at specific wavelengths.  
%A coral reef is known to produce fluorescence emissions on Earth.
Biofluorescence, similar to that shown by coral reefs on Earth, has been suggested as a new potential biosignature for exoplanets experiencing strong UV radiation from F stars \citep{o2018biofluorescent} and M stars \citep{o2019biofluorescent}.
It might work if the fluorescence were emitted very efficiently according to gained photons in their habitats.
As mentioned above, photosynthetic pigments are a potential emitter of biofluorescence. However, the yield and detectability of photosynthetic fluorescence on the surface of exoplanets have not yet been examined.

Finding surface biosignatures on Earth-like exoplanets, including the potential detectability of biofluorescence, would be
one of the important goals of future astronomy and may become possible
with future space missions such as
the Large UV/Optical/IR Surveyor (LUVOIR) or the Habitable Exoplanet Observatory (HabEx),
and next-generation extremely large ground-based telescopes (TMT, ELT, and GMT) observing in reflected light.
Thus, it is important to quantitatively evaluate the detectability of
any potential surface biosignature using expected specifications of specific future missions.

This study made the first attempt to investigate the detectability of photosynthetic fluorescence on Earth-like exoplanets. The remainder of this paper is structured as follows:
Section 2 describes the surface vegetation model for an Earth-like planet in the habitable zone and fluorescence emissions based on the photoresponse of photosynthetic organisms.
Section 3 shows the expected fluorescence emissions in the reflected light spectra on an Earth-like planet around an M dwarf or the Sun.
In Section 4, we discuss the physiological conditions of photosynthesis that enhance fluorescence emissions and its unique features for future detection, including false-positive signals and seasonal changes. Additionally, we present the detectability of biofluorescence by a future space-based telescope assuming the LUVOIR telescope parameters, and the key spectral feature possibly useful for the detection by follow-up observations with high-dispersion spectroscopy.
In the last section, we summarize our paper.

\section{Materials and Methods}
\label{desc:methods}

We assume that the radiation from a planetary surface is the sum of the reflected light on the surface and the fluorescence emission from photosynthesis. The outgoing flux on the surface and at the top of atmosphere (TOA) is given by:
\begin{eqnarray}
  \label{eqn:flux_general}
  F^{\uparrow}_\mathrm{surface}(\lambda) &=& F^{\downarrow}_\mathrm{surface}(\lambda) R(\lambda)+{F_\mathrm{fluor.}}(\lambda),\\
  F^{\uparrow}_\mathrm{TOA}(\lambda) &=& F^{\uparrow}_\mathrm{surface} (\lambda) T(\lambda),
\end{eqnarray}
where $\lambda$ is the wavelength, $T (\lambda)$ is the atmospheric transmittance (see Section \ref{desc:st-rad}), $R(\lambda)$ is the surface reflectance of a planet (see Section \ref{desc:refl}),
$F^{\uparrow}_\mathrm{surface}(\lambda)$ is the upward flux from a planetary surface, $F^{\uparrow}_\mathrm{TOA}(\lambda)$ is the reflected flux at the TOA, and ${F_\mathrm{fluor.}}(\lambda)$ is the net fluorescence emission from photosynthesis.
$F^{\downarrow}_\mathrm{surface}(\lambda) = F^{\downarrow}_\mathrm{TOA} (\lambda) T (\lambda)$ is the downward flux from the planetary atmosphere to the surface, where $F^{\downarrow}_\mathrm{TOA}$ is the incident flux from a host star at the TOA (see Figure \ref{fig:stars} and Section \ref{desc:st-rad} below).
We neglect the effects of thermal emission in all the cases and Rayleigh scattering in most cases, as both processes contribute little radiation to our spectral region of interest (600-1000 nm).
%Since black body radiation contributes little to our spectral region of interest, we neglected black body radiation in this study.
The transmittance $T (\lambda)$ in the atmosphere of an Earth-like planet through geological evolution was obtained from \cite{rugheimer2018spectra}, which was calculated by a 1D coupled radiative/convective-photochemical model for a planetary atmosphere~\citep[see also][]{pavlov2002mass,kasting1986climatic,segura2005biosignatures}.

\subsection{Stellar Radiation}
\label{desc:st-rad}

\begin{figure}
 \begin{center}
 \includegraphics[width=180mm]{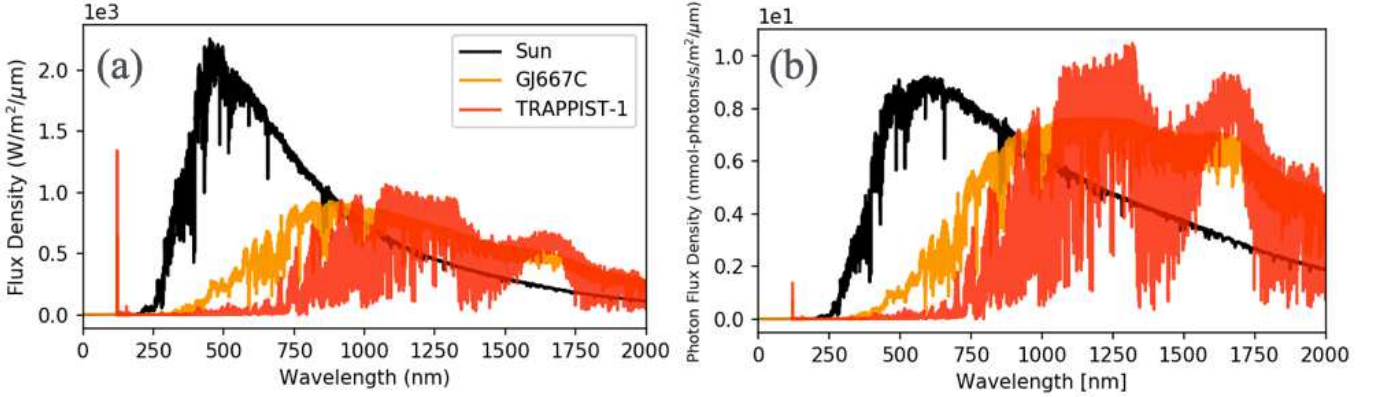}
 \end{center}
\caption{(a) Incident radiation and (b) photon flux density at the top of atmosphere (TOA) of an Earth-like planet around the Sun, GJ667C, and TRAPPIST-1. The spectral data of the Sun, GJ667C, and TRAPPIST-1 were obtained from \citet{meftah2018solar}, \citet{france2016muscles}, and \citet{lincowski2018evolved}, respectively.
}
 \label{fig:stars}
\end{figure}

Two nearby M dwarfs, GJ667\,C and TRAPPIST-1, have candidate planets in a habitable zone (HZ). 
We considered fluorescence emissions from photosynthesis on an Earth-like planet in an HZ around GJ667\,C, TRAPPIST-1, and the Sun.
We extracted the incident stellar flux from high-resolution spectral data for the Sun \citep{meftah2018solar}, GJ667\,C \citep{france2016muscles}, and TRAPPIST-1 \citep{lincowski2018evolved}.
The incident flux $F^{\downarrow}_\mathrm{TOA}$ received by an Earth-like planet around GJ667\,C, and TRAPPIST-1 is scaled by the current location of GJ667C\,c, and TRAPPIST-1e. GJ667\,C, and TRAPPIST-1 are modeled as M1V and M8V stars.
Figure~\ref{fig:stars} shows the incident flux received by an Earth-like planet at the TOA around the Sun, GJ667\,C, and TRAPPIST-1.

%%%%%%%%%%%%%%%%%%%%%%%%%%%%%%%%%%%%%%%%%%
%%%%%%%%%%%%%%%%%%%%%%%%%%%%%%%%%%%%%%%%%%

\subsection{Fluorescence from Photosynthesis}
\label{desc:fl}

\begin{figure}
 \begin{center}
 \includegraphics[width=90mm]{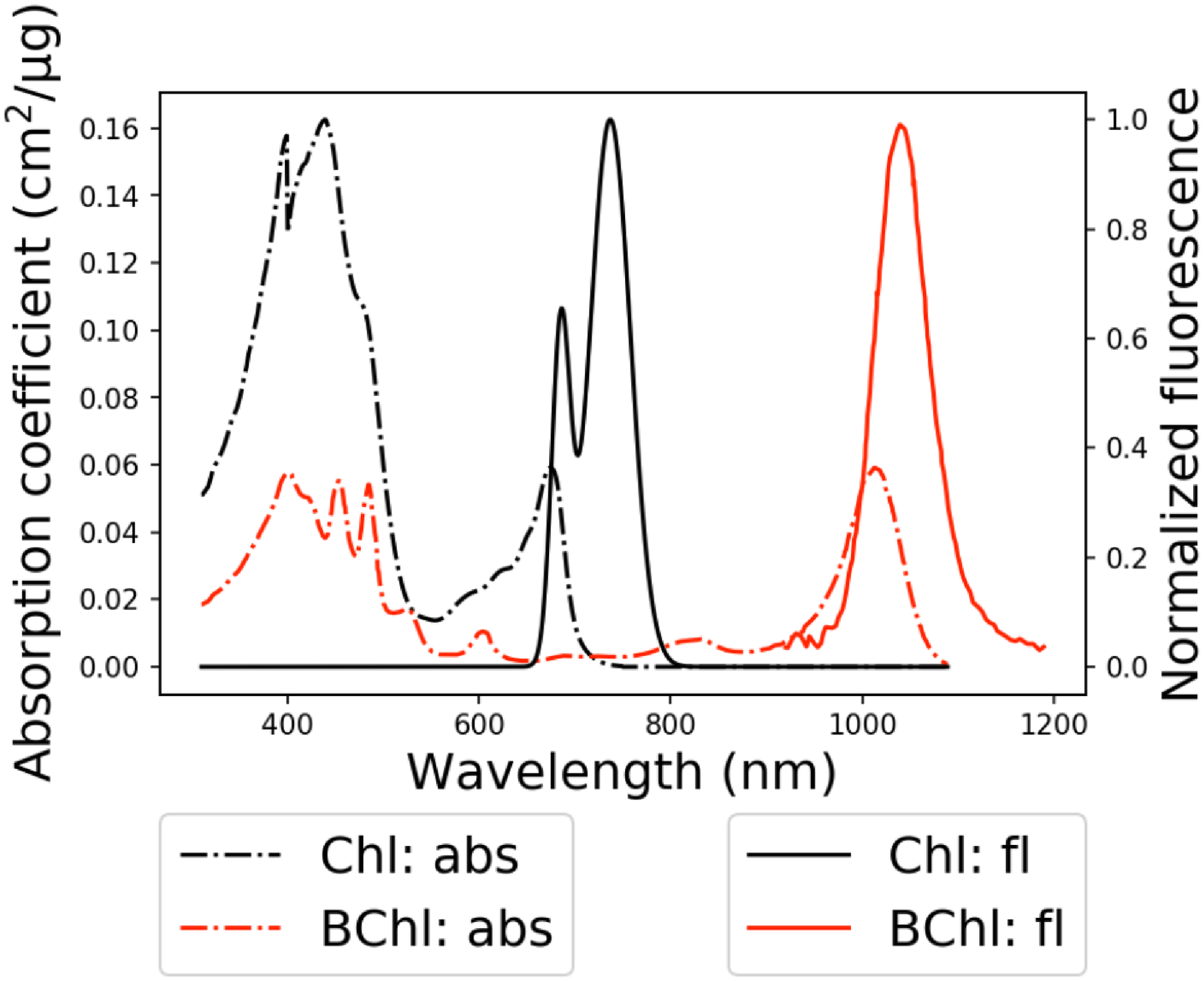}
 \end{center}
% \vspace*{1.0mm}
\caption{Fluorescence ($f(\lambda)$: solid curves) and photoabsorption ($\sigma (\lambda)$: dashed curves) spectra for chlorophylls (Chl: black) and bacteriochlorophylls (BChl: red). The absorption coefficient of chlorophylls in units of cm$^2$\,$\mu$g$^{-1}$ was obtained from~\cite{feret2008prospect}. The fluorescence spectrum is expressed by the Gaussian functions given in~\cite{frankenberg2012remote} and \cite{guanter2010developments}. 
For bacteriochlorophylls, $f(\lambda)$ and $\sigma (\lambda)$ adopt those of the LH1--RC complex of a bacteriochlorophyll $b$ containing purple photosynthetic bacteria~\citep{magdaong2016carotenoid}. The nondimensional absorption spectrum for BChl is normalized at the peak value in the longest absorption band, the Q band, of the Chl. Two fluorescence spectra are normalized at their peak values.
}
 \label{fig:abs}
%\vspace*{-1mm}
\end{figure}

Fluorescence emissions from a planetary surface ${F_\mathrm{fluor.}^{'}}$ are expressed as:
\begin{equation}
  \label{eqn:unscaled}
  {F_\mathrm{fluor.}^{'}} (\lambda) = s c_v \pi {F_\mathrm{fluor.}^\mathrm{std}} \times f(\lambda),
\end{equation}
where $c_v$ is the surface coverage of vegetation (see Section~\ref{desc:refl}), $s$ is the scaling factor from the standard observed fluorescence emission reflecting the photosynthetic activity,
and ${F_\mathrm{fluor.}^\mathrm{std}}$ is the standard fluorescence intensity from vegetation based on field measurements.
The spectral shape of fluorescence emissions from a photosynthetic organism at wavelength $\lambda$ is defined by $f(\lambda)$. 
In this study, $F_\mathrm{fluor.}^\mathrm{std} = 1.0$ (W\,m$^{-2}$\,$\mu\mathrm{m}^{-1}$\,sr$^{-1}$) \citep{du2019sifspec,yao2021new} and
$s = 0, 1, 5,\,\mathrm{and}\,10$. 
The net fluorescence intensity ${F_\mathrm{fluor.}}$ is calculated by considering acquired photons at the habitat using ${F_\mathrm{fluor.}^{'}}$ in Equation~(\ref{eqn:unscaled}) as:
\begin{equation}
  \label{eqn:fs}
  {F_\mathrm{fluor.}} (\lambda) = \frac{\chi}{\chi_0}{F_\mathrm{fluor.}^{'}} (\lambda), 
\end{equation}
\begin{equation}
  \label{eqn:chi}
  \chi \equiv \int n (\lambda) \sigma (\lambda) d \lambda, 
\end{equation}
\begin{equation}
    \label{eqn:chi0}
  \chi_{0} \equiv \int n_\mathrm{sun, ref.} (\lambda) \sigma_\mathrm{chls} (\lambda) d \lambda,
\end{equation}
where $\chi$ is the light absorption efficiency, $n (\lambda)$ is the photon flux density at the planetary surface, and $\sigma (\lambda)$ is the absorption coefficient of a photosynthetic pigment.
$\chi_{0}$ represents the standard absorption efficiency on Earth.
The subscript chls on $\sigma (\lambda)$ represents chlorophylls (see Chl:abs in Figure~\ref{fig:abs}).
$n_\mathrm{sun, ref.} (\lambda)$ is the photon flux density on the surface of the Earth from the reference solar spectral irradiance at an air mass of 1.5 (National Renewable Energy Laboratory), which corresponds to a typical irradiance for Earth vegetation.
%~\citep{astm2012standard}. 

We considered an incident flux from a star under two sky conditions, a clear sky and 60\% cloud cover, to estimate the reflectance at the TOA and $\chi$ on the ground, in accordance with the setup for the simulation.
We assumed the clear sky condition, if not specified, and the cloud condition appeared only in Section~\ref{desc:luvoir} (Figure~\ref{fig:ref_bchl_cloud}). 
In the cloudy condition, 60\% of the radiation is reflected in three kinds of clouds, and 40\% of the radiation reaches the ground.
For the clouds, we assumed that 40\% are low water clouds, 40\% are high water clouds, and 20\% are high ice clouds~\citep{gao2003water} at 1, 6, and 12\,km altitude, respectively.
To model Earth-like conditions, 
the effect of Rayleigh scattering in a planetary atmosphere was also considered in the cloudy condition using a previously described empirical approach by \cite{bucholtz1995rayleigh} (see Appendix A for more details).

Equation (\ref{eqn:fs}) indicates that ${F_\mathrm{fluor.}^{'}}$ is linearly scaled to the number of incoming photons that are absorbed by chlorophylls at the planetary surface.
In other words, chlorophylls can emit strong fluorescence if the spectral shapes of $n (\lambda)$ and $\sigma (\lambda)$ match well.
Note that $n (\lambda)$ is exactly the same as $F^{\downarrow}_\mathrm{surface} (\lambda)$, and its unit is shown in Figure~\ref{fig:stars}(b) ($F^{\downarrow}_\mathrm{TOA}$ in the figure).
This treatment in Equations~(\ref{eqn:unscaled}) and~(\ref{eqn:fs}) can be applied to the relationship between the incoming photons and the photons emitted as fluorescence on an Earth-like planet around various stars other than the Sun.

Figure~\ref{fig:abs} shows the normalized spectra of fluorescence $f(\lambda)$ and absorption coefficient $\sigma (\lambda)$ for Chls and BChls. The peak wavelength of $f(\lambda)$ is red-shifted from that of $\sigma (\lambda)$, which is called the Stokes shift~\citep{lakowicz2006principles}.
There are two absorption bands in the $\sigma (\lambda)$ of chlorophylls: the B band (known as the Soret band) in the short-wavelength region and the Q band in the long-wavelength region.
The primary fluorescence emission is derived from the Q absorption band. 
In this study, we modeled $f(\lambda)$ for Chls as the superposition of two Gaussian distributions~\citep{frankenberg2012remote,guanter2010developments} with means of 680\,nm (PSII) and 740\,nm (PSI and PSII).
$\sigma (\lambda)$ for Chl uses the model vegetation with chlorophylls ($\sigma_\mathrm{chls} (\lambda)$)~\citep{feret2008prospect}. 
We obtained $f(\lambda)$ and $\sigma (\lambda)$ for BChls from the spectral data for the LH1--RC complex, the supramolecular complex of the light-harvesting core antenna (LH1), and the RC in a bacteriochlorophyll $b$ containing purple photosynthetic bacteria \citep[see Figure 3 in ][]{magdaong2016carotenoid}).
Note that we used only $\sigma (\lambda)$ in the Q band for calculating $\chi$ and $\chi_{0}$ because free Chls and BChl--protein complexes in each solution affect each spectrum in the B band to different degrees.

%%%%%%%%%%%%%%%%%%%%%%%%%%%%%%%%%%%%%%%%%%
%%%%%%%%%%%%%%%%%%%%%%%%%%%%%%%%%%%%%%%%%%

\subsection{Surface Vegetation}
\label{desc:refl}

\begin{figure}
 \begin{center}
 \includegraphics[width=80mm]{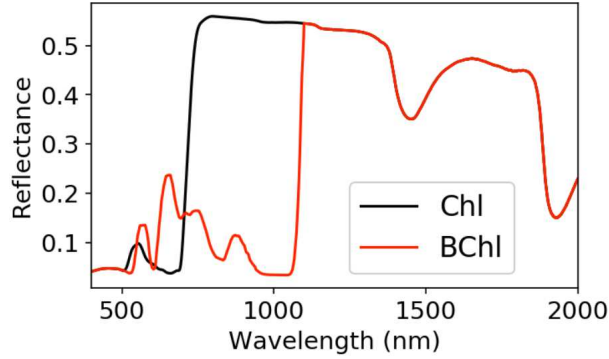}
 \end{center}
%  \vspace*{5.0mm}
\caption{The reflectance of vegetation estimated from radiation transfer calculations for two leaf models: Chl (``standard'') and BChl (``hypothetical'') ~\citep{jacquemoud1990prospect,feret2008prospect}. The light absorption spectrum for Chls and BChls uses $\sigma (\lambda)$ in Figure~\ref{fig:abs}.}
 \label{fig:exoveg}
%\vspace*{-3mm}
\end{figure}

To determine the detectability of vegetation fluorescence, we use two leaf models for our experiments: one which assumes the reflectance spectrum and fluorescence of standard chlorophyll and another that uses a scaled version of the spectrum of bacteriochlorophyll.
The reflectance of a planet is expressed as $R(\lambda) =\sum_{i}c_i r_{i} (\lambda)$, where $i$ denotes the surface type, $c_i$ is the fraction of the surface coverage of type $i$, and $r_{i}$ the reflectance of type $i$.
We obtained the reflection spectra for various surface types including vegetation, ocean, and coast from the USGS Digital Spectral Library and the ASTER Spectral Library \citep{BALDRIDGE2009711}. The detailed compositions used in this paper are summarized in Table~\ref{table:settings}.
The reflectance of the surface vegetation $r_v$ is estimated from radiation transfer calculations for a modeled leaf~\citep{jacquemoud1990prospect,feret2008prospect}, using $\sigma (\lambda)$ over all the wavelengths shown in Figure~\ref{fig:abs}.
Figure~\ref{fig:exoveg} shows the reflectance of a Chl-based leaf (``standard'') and a BChl-based leaf (``hypothetical'').
In the latter case, we assumed the vegetation on a different planet has a photosynthetic pigment whose optical property is the same as BChl exhibiting the VRE in the longer wavelength region as shown in Figure~\ref{fig:exoveg}.
As the input to the radiative transfer calculations, we used the absorption spectra of Chl~\citep{feret2008prospect} and BChl~\citep{magdaong2016carotenoid}.
The unitless absorption spectrum for BChl is normalized at the peak in that for Chl, unlike the calculations of $\chi$ and $\chi_0$ in Section~\ref{desc:fl}.
As shown in Figure~\ref{fig:exoveg}, both Chl- and BChl-based leaves show a large reflectance (i.e., the VRE) in the wavelength ranges around 700--750 and 1000--1100\,nm.
The green bump around 500\,nm is observed in the reflectance for Chl, and larger and broader bumps are observed from $\sim$500 to 950\,nm for BChls by the larger difference in the wavelength between the B and Q bands than observed for Chl.
Like many kinds of photosynthetic organisms, the organisms with BChl could have acquired accessory pigments such as carotenoids (Cars) that absorb photons with wavelengths between the B and Q bands of Chl~\citep{cogdell1978carotenoids}.
The effective light absorption by accessory pigments can suppress the increase in reflectance.
With or without accessory pigments, the bump for BChl does not affect fluorescence emissions in the wavelength (see Figures~\ref{fig:ref_bchl},~\ref{fig:ref_bchl_2}, and ~\ref{fig:39}). 
The low reflectance from $\sim$500 to 700\,nm (1000\,nm), due to the light absorption by Chls (BChls), affects the reflectance of the planet. The degree of reduction in the overall planetary reflectance varies depending on the surface coverage by vegetation.

%%%%%%%%%%%%%%%%%%%%%%%%%%%%%%%%%%%%%%%%%%
%%%%%%%%%%%%%%%%%%%%%%%%%%%%%%%%%%%%%%%%%%

\section{Results}
\label{desc:results}

We considered three fluorescence cases on an Earth-like planet at different stages of atmospheric evolution around the Sun, GJ667\,C, and TRAPPIST-1 for different surface biosignatures: Earth-like (Chl) vegetation, hypothetical BChl-based vegetation, and biological fluorescence without any surface vegetation.
Our models for the surface compositions, vegetation, fluorescence types, and atmospheric compositions, i.e., transmittance, are summarized in Table$~\ref{table:settings}$. Mod-earth corresponds to the surface condition for the Modern Earth, leading to a lesser contribution of fluorescence emissions than in the other two cases.
The veg-only models are considered optimistic conditions for fluorescence emissions where vegetation covers the whole planetary surface. The veg-land models, with 70 \% ocean, 2\% coast, and 28\% land covered with the vegetation, lie between the mod-earth and veg-only models. 
As mentioned in Section~\ref{desc:methods}, we considered two leaf models for land vegetation: Chl-based vegetation
and BChl-based vegetation.
For the atmospheric compositions of an Earth-like planet, we adopted the Modern Earth model at 0.0\,Ga (oxygen-rich atmosphere), the Paleoproterozoic Earth model at 2.0\,Ga (oxygen-poor atmosphere), and the Archean Earth model at 3.9\,Ga (anoxic atmosphere) \citep[see Table 1 in][]{rugheimer2018spectra}.
As an extreme case, we assumed the presence of photosynthetic bacteria with BChl spread over the land and ocean on an Archean-Earth-like planet with no surface vegetation.
We assumed a clear sky for all atmospheric conditions in Section~\ref{desc:results}.

%%%%%%%%%%%%%%%%%%%%%%%%%%%%%%% table for all setting %%%%%%%%%%%%%%%%%%%%%%%%%%%%%%% 

\begin{table}[h]
\centering
\hspace{-2.5cm}
\begin{tabular}{|c||c|c|c |c|c|} \hline
  Model name & Surface compositions & Surface vegetation & Fluorescence type & $T(\lambda)$ & $c_v$ \\ \hline \hline
   veg-only 0C & & Chl surf. & Chl fluor. & 0.0 Ga &  \\ \cline{1-1}\cline{5-5}
    veg-only 2C & 100\% vegetation & & & 2.0 Ga &  1.00\\ \cline{1-1}\cline{3-5}
   veg-only 0B & & BChl surf. & BChl fluor. & 0.0 Ga &  \\ \cline{1-1}\cline{5-5}
    veg-only 2B & & & & 2.0 Ga &  \\ \cline{1-6}
   veg-land 0C & & Chl surf. & Chl fluor. & 0.0 Ga &  \\ \cline{1-1}\cline{5-5}
    veg-land 2C & 70\% ocean, 2\% coast & & & 2.0 Ga &  0.28\\ \cline{1-1}\cline{3-5}
   veg-land 0B & and 28\% vegetation & BChl surf. & BChl fluor. & 0.0 Ga &  \\ \cline{1-1}\cline{5-5}
    veg-land 2B & & & & 2.0 Ga &  \\ \cline{1-6}
   mod-earth 0C & & Chl surf. & Chl fluor. & 0.0 Ga &  \\ \cline{1-1}\cline{5-5}
    mod-earth 2C &70\% ocean, 2\% coast & & & 2.0 Ga &   0.168\\ \cline{1-1}\cline{3-5}
   mod-earth 0B & and 28 \% mixed land  & BChl surf. & BChl fluor. & 0.0 Ga &  \\ \cline{1-1}\cline{5-5}
    mod-earth 2B & (incl. 16.8\% vegetation) & & & 2.0 Ga &  \\ \cline{1-6}
   anoxic B & 70\% ocean, 2\% coast and & - & BChl fluor. & 3.9 Ga & 0.72 \\ 
    & 28\% mixed land at 3.9 Ga & & & &  \\ \cline{1-6}
\end{tabular}
\caption{Surface composition, vegetation, its fluorescence types, and atmospheric transmittance ($T(\lambda)$) for all the cases in this paper. 
 Mixed land is composed of 60\% vegetation (16.8\% in total), 15\% snow, 9\% granite, 9\% basalt, and 7\% sand \citep{BALDRIDGE2009711}; mixed land at 3.9 Ga means the land model of the Archean Earth at 3.9\,Ga, which is composed of 35\% basalt, 40\% granite, 15\% snow, and 10\% sand. Chl surf. and BChl surf. correspond to reflection spectra of Chl and BChl in Figure \ref{fig:exoveg}, respectively. 
 The spectral shapes of fluorescence emissions $f(\lambda)$ for Chl fluor. and BChl fluor. correspond to the fluorescence spectra of Chl and BChl in Figure~\ref{fig:abs}, respectively; their intensities $F_\mathrm{flour.}$ are scaled in Equations (\ref{eqn:unscaled}) and (\ref{eqn:fs}).
  $c_v$ is given by the relationship between the surface coverage of vegetation and the fluorescence emission. $s$=\{0, 1.0, 5.0, 10.0\}. We obtained $T(\lambda)$ at 0.0, 2.0, and 3.9\,Ga from~\citet{rugheimer2018spectra}.\label{table:settings}}
\end{table}

%%%%%%%%%%%%%%%%%%%%%%%%%%%%%%%%%%%%%%%%%%%%%%%%%%%%%%%%%%%%%%%%%%%%%%%%%%%%%%%%%%%%%%%%%%%%%

\subsection{Case-1: Planets with Earth-Like Vegetation}

\begin{figure}
 \begin{center}
 \includegraphics[width=180mm]{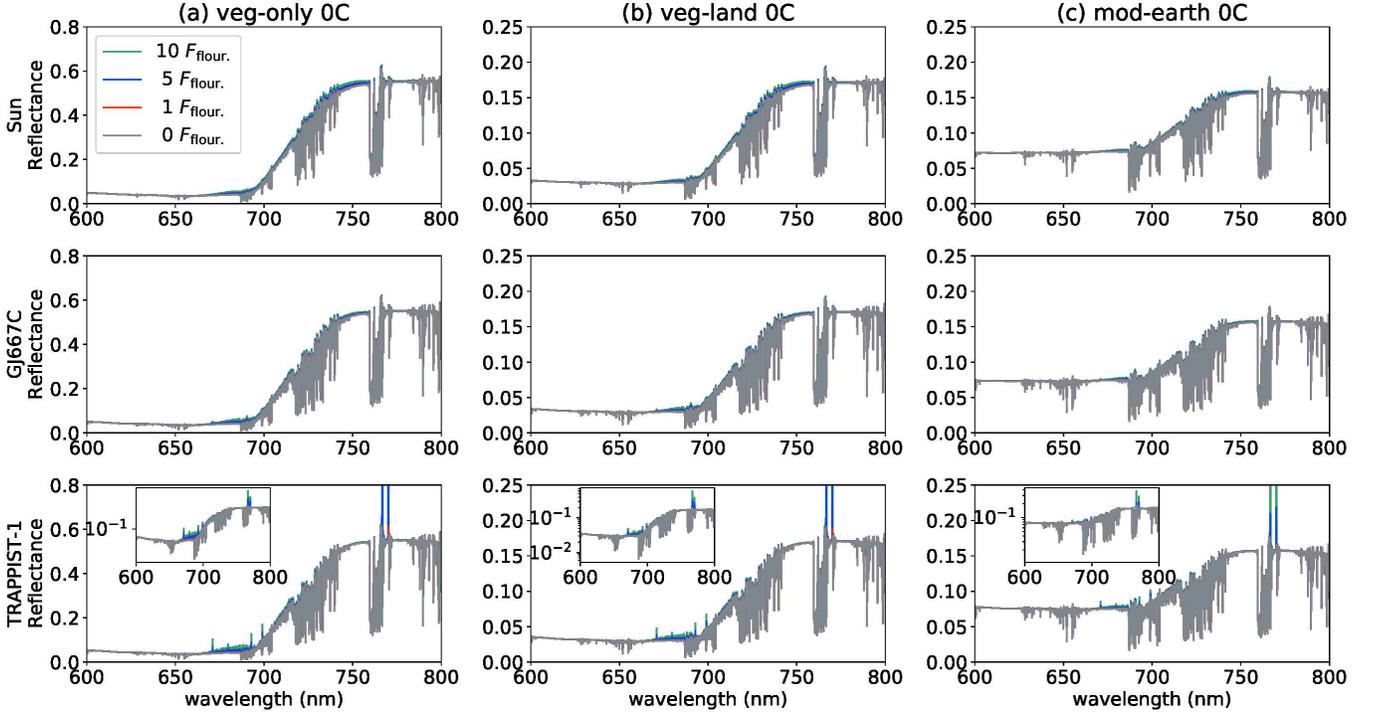}
 \end{center}
\caption{Reflectance of an Earth-like planet with the Modern Earth's atmosphere (0.0Ga) around the Sun, GJ667C, and TRAPPIST-1.
The three colors represent the reflected light from a planet with $F_\mathrm{fluor.}$ ($s = 1$: red), $5F_\mathrm{fluor.}$ ($s = 5$: blue), and $10F_\mathrm{fluor.}$ ($s = 10$: green), where $F_\mathrm{fluor.}$ is the fluorescence emission from chlorophylls observed on the Earth. No-fluorescence emission models are also indicated by gray lines.
We assumed Earth-like vegetation (chlorophylls) covers the planetary surface (see Table 1 for model details). The reflectance is defined here as $F^{\uparrow}_\mathrm{TOA} (\lambda)$/$F^{\downarrow}_\mathrm{TOA} (\lambda)$, where $F^{\uparrow}_\mathrm{TOA} (\lambda)$ is the light reflected from the ground at the top of atmosphere (TOA), and $F^{\downarrow}_\mathrm{TOA} (\lambda)$ is the flux at TOA induced by stars. For each case around TRAPPIST-1, the reflectance with a logarithmic scale is also shown as the inset plot.
}
 \label{fig:ref_chl}
\end{figure}

\begin{figure}
 \begin{center}
 \includegraphics[width=180mm]{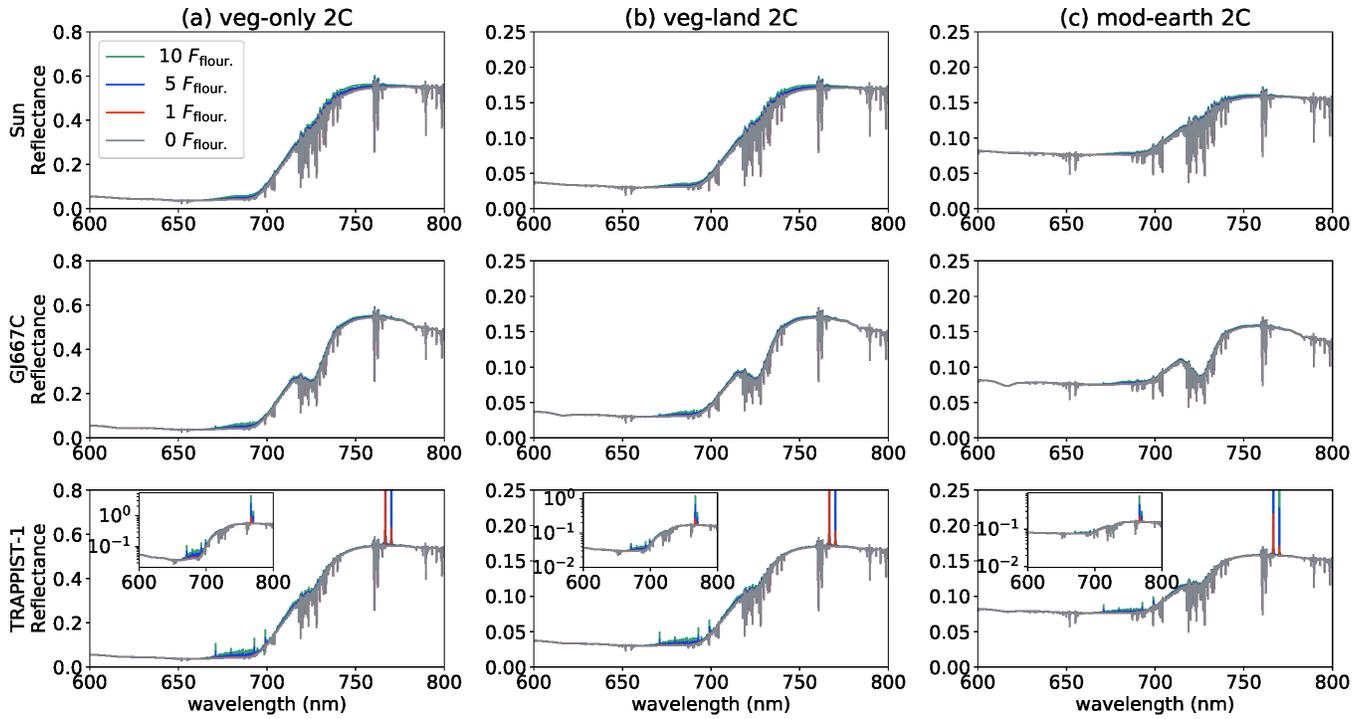}
 \end{center}
\caption{The same as Figure~\ref{fig:ref_chl}, but for an Earth-like planet with an oxygen-poor atmosphere (2.0 Ga).}
 \label{fig:ref_chl_2}
\end{figure}

In case-1, Earth-like vegetation (Chl) emits fluorescence on the surface of an Earth-like planet.
The fluorescence emissions from chlorophyll are visible at the wavelengths from 650 to 800\,nm, as shown in Figure~\ref{fig:abs}.
To determine the contribution of fluorescence from planets, 
the reflectance is defined as $F^{\uparrow}_\mathrm{TOA} (\lambda)$/$F^{\downarrow}_\mathrm{TOA} (\lambda)$ and calculated.
Figures~\ref{fig:ref_chl} and~\ref{fig:ref_chl_2} show the reflectance of an Earth-like planet with the Modern Earth's atmosphere (0.0\,Ga) and an oxygen-poor atmosphere (2.0\,Ga), respectively.
The O$_2$, O$_3$, CH$_4$, and H$_2$O absorption features in the atmosphere are imprinted in the reflectivity in the visible--NIR wavelengths from 600--800\,nm. 
The oxygen-poor-atmosphere models show less conspicuous patterns in the reflectance profile in the 700 to 750\,nm wavelength region.
The reflectivity between 600 and 700\,nm is nearly constant but increases with decreasing surface coverage of vegetation.
The VRE is observed as the steep rise in the reflectance from 700 to 750\,nm (also see Figure~\ref{fig:exoveg}), whereas the reflectance excess due to fluorescence is quite small, even in optimistic conditions (veg-only models).
Note that the red curve with $1 F_\mathrm{fluor.}$ around the Sun in the mod-earth model (Figure~\ref{fig:ref_chl}), corresponding to the modern earth fluorescence, is hardly seen.
Around TRAPPIST-1, however, sharp increase in the reflectance around 770\,nm is due to the strong absorption of potassium in the stellar atmosphere.
As a result, we observed similar features in the light reflected from an Earth-like planet with different atmospheric compositions around TRAPPIST-1
%.
%The reflectance excess around 770\,nm would {\com provide a clue to biofluorescence from an Earth-like planet around an ultracool red dwarf}
 (see Section~\ref{desc:appa} for further discussion).

\begin{figure}
 \begin{center}
 \includegraphics[width=180mm]{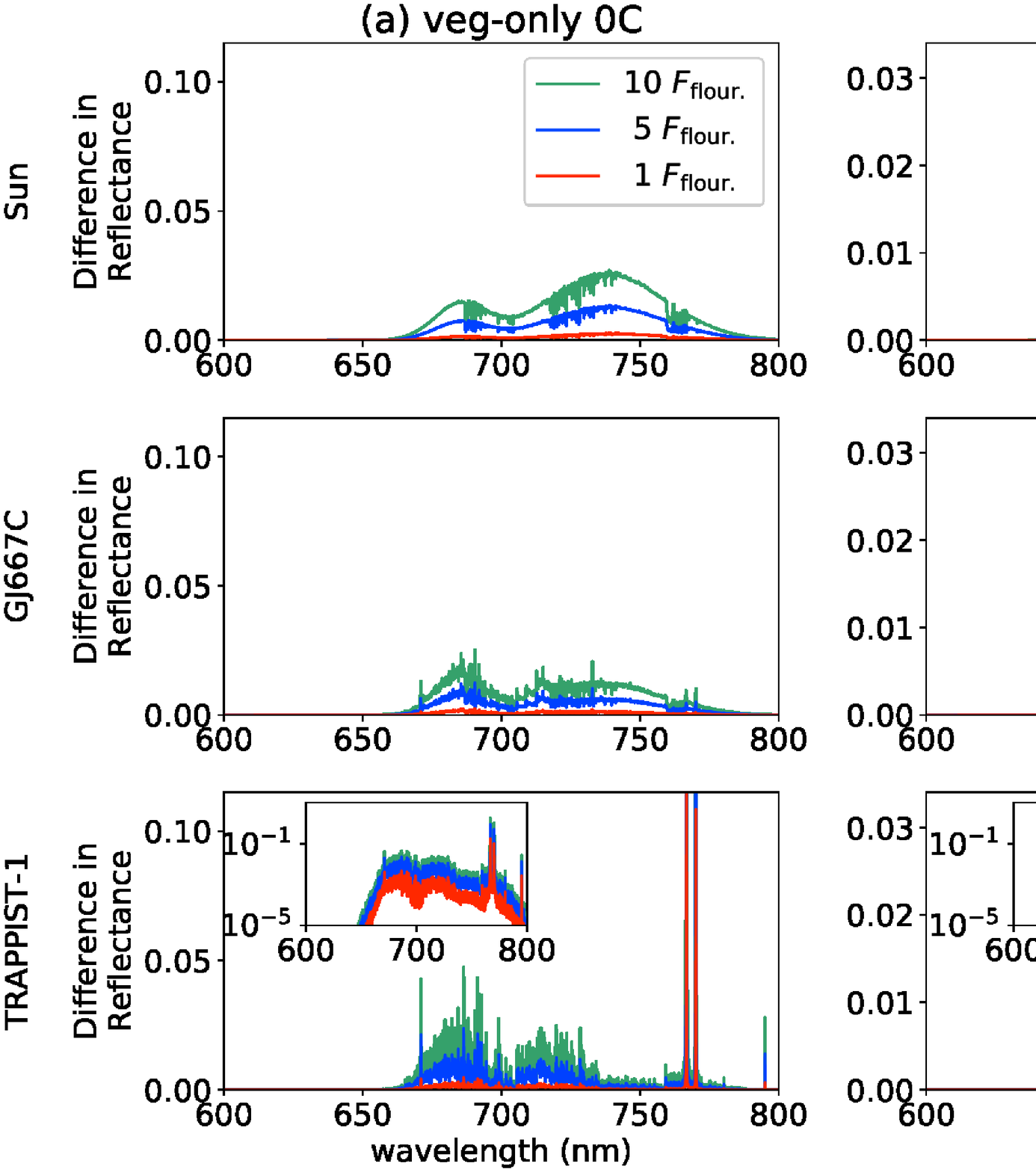}
 \end{center}
\caption{Reflectance excess due to chlorophyll fluorescence emissions on an Earth-like planet with the Modern Earth's atmosphere.}
 \label{fig:diff_chl}
\end{figure}

Figure~\ref{fig:diff_chl} shows the reflectance excess due to fluorescence emissions on an Earth-like planet with the Modern Earth's atmosphere.
Atmospheric absorptions, such as H$_2$O, O$_2$, and O$_3$, weaken the Gaussian features in the fluorescence emissions from an Earth-like planet around the Sun.
The fluorescence from chlorophylls around 740\,nm is less pronounced for a planet around M dwarfs than one around the Sun because of weaker radiation flux in the wavelength region of 700--750\,nm (see Figure\,\ref{fig:stars}).
In addition, a sudden increase in reflectance due to the VRE obscures the fluorescence emission around 740\,nm (see Figures~\ref{fig:ref_chl} and~\ref{fig:ref_chl_2}).
As a result, the Chl fluorescence around 680\,nm emitted from PSII on an Earth-like planet would be the most promising feature for detection (see Figure~\ref{fig:abs}).
Note that nonphotochemical quenching processes can decrease the fluorescence intensity around 680\,nm, and the fluorescence emission is further reduced by the reabsorption of photons within the canopy~\citep{porcar2021chlorophyll}.

%%%%%%%%%%%%%%%%%%%%%%%%%%%%%%%%%%%%%%%%%%
%%%%%%%%%%%%%%%%%%%%%%%%%%%%%%%%%%%%%%%%%%

%%%%%%%%%%%%%%%%%%%%%%%%%%%%%%%%%%%%%%%%%%
%%%%%%%%%%%%%%%%%%%%%%%%%%%%%%%%%%%%%%%%%%
 
%%%%%%%%%%%%%%%%%%%%%%%%%%%%%%%%%%%%%%%%%%
%%%%%%%%%%%%%%%%%%%%%%%%%%%%%%%%%%%%%%%%%% 

\subsection{Case-2: Planets with Bacteriochlorophylls-Based Vegetation}

\begin{figure}
 \begin{center}
 \includegraphics[width=180mm]{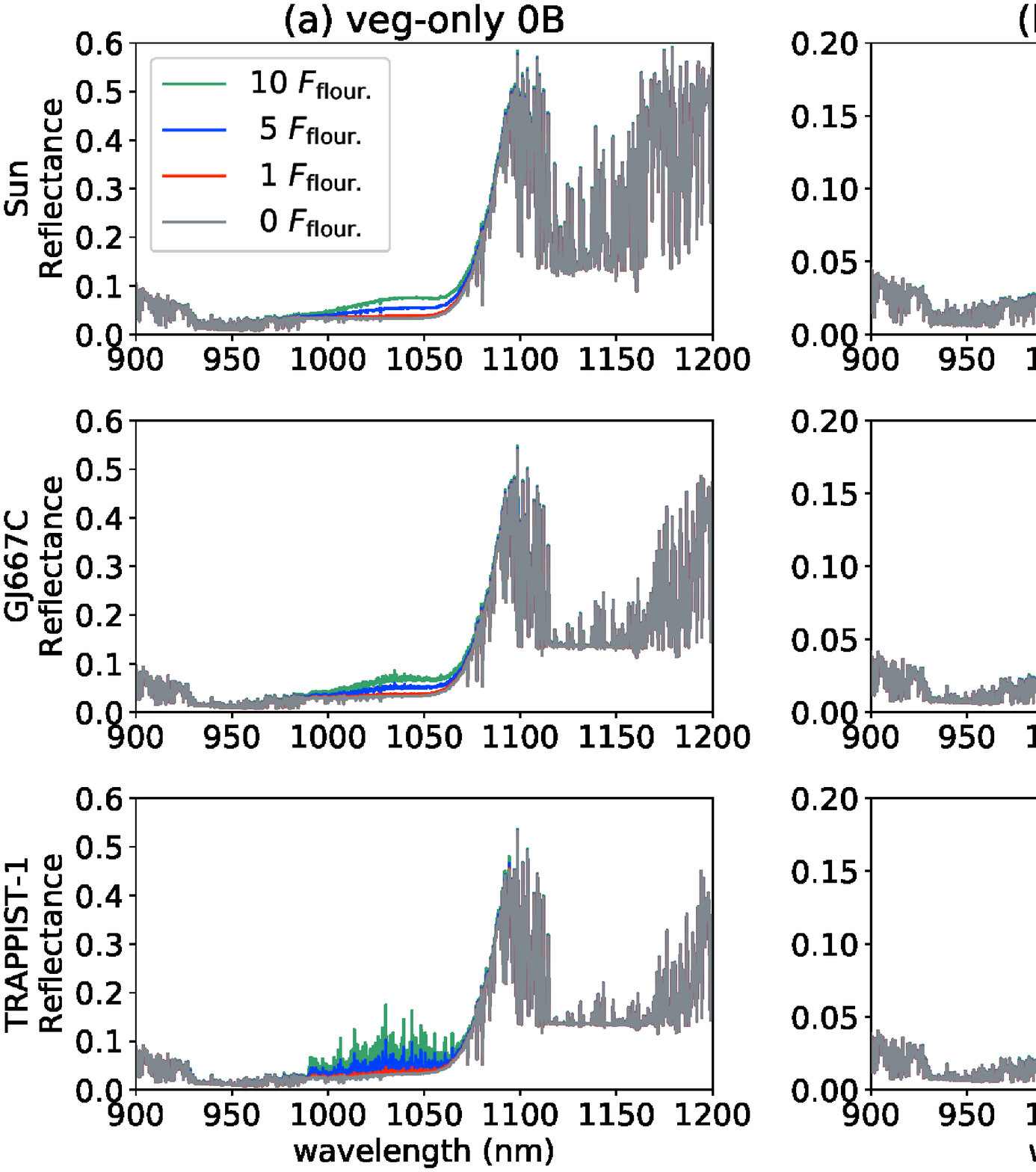}
 \end{center}
\caption{The same as Figure\,\ref{fig:ref_chl} but for the reflectance of a planet covered with bacteriochlorophyll-based vegetation.}
 \label{fig:ref_bchl}
\end{figure}

\begin{figure}
 \begin{center}
 \includegraphics[width=180mm]{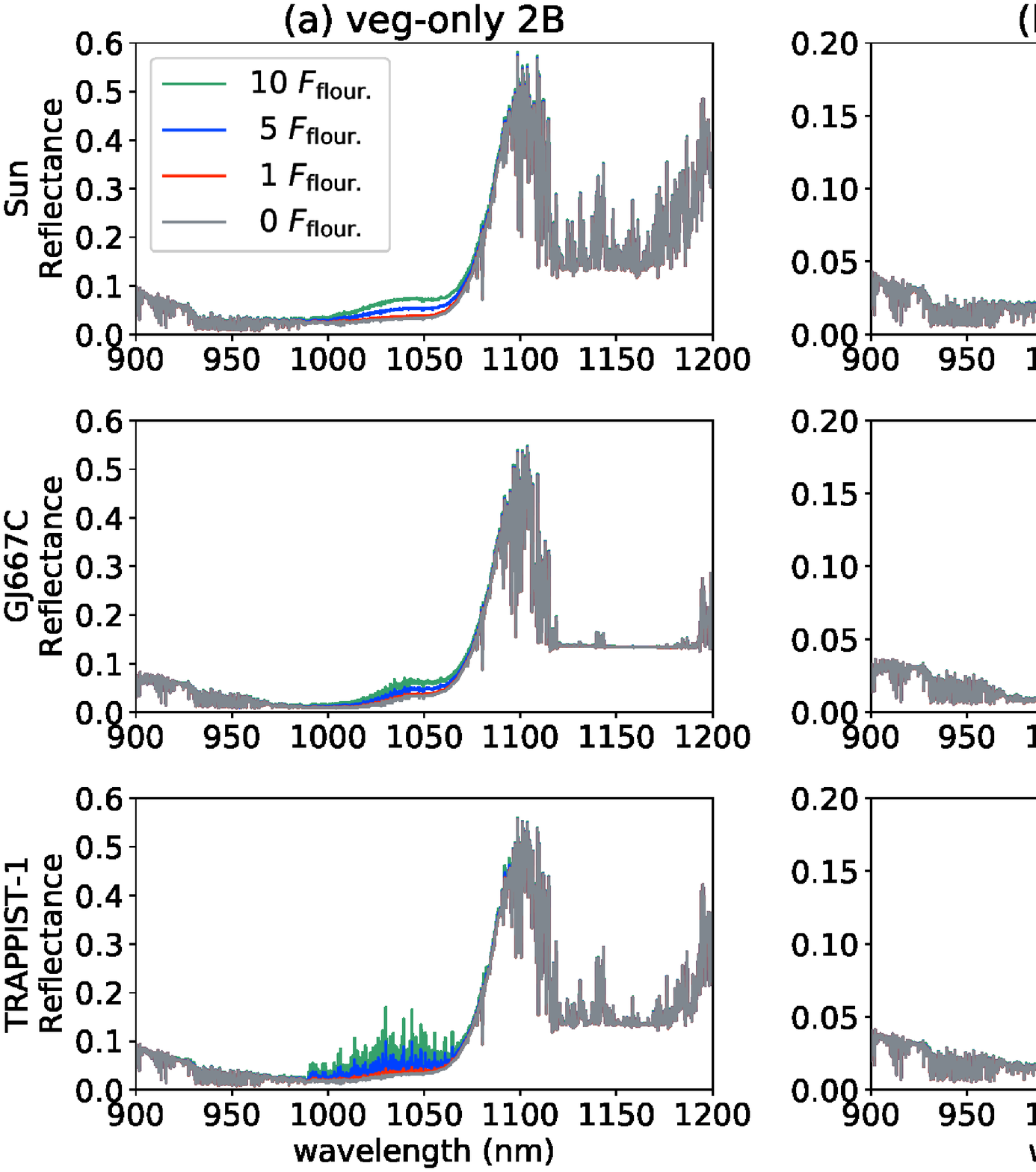}
 \end{center}
\caption{The same as Figure~\ref{fig:ref_bchl}, but for a planet with an oxygen-poor atmosphere (2.0 Ga).}
 \label{fig:ref_bchl_2}
\end{figure}

\begin{figure}
 \begin{center}
 \includegraphics[width=180mm]{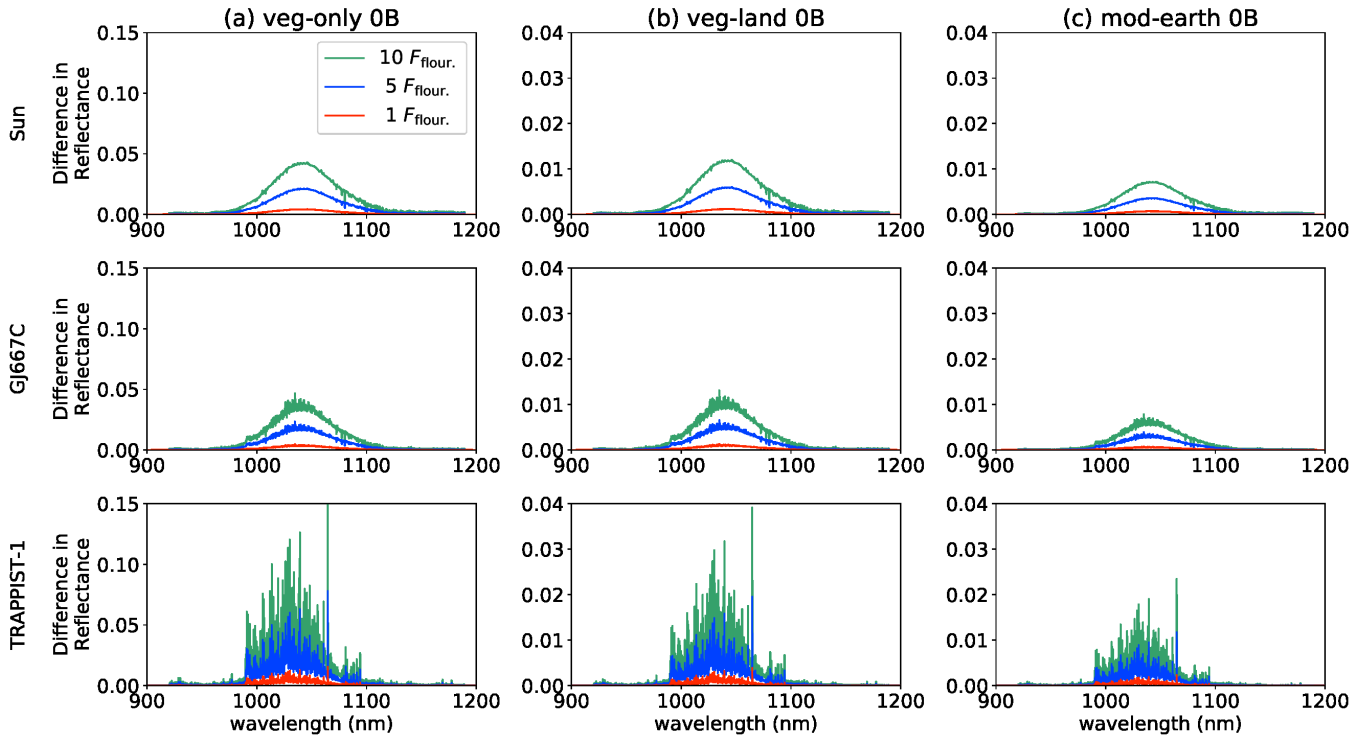}
 \end{center}
\caption{Reflectance excess due to bacteriochlorophyll fluorescence emissions on an Earth-like planet with the Modern Earth's atmosphere.}
 \label{fig:diff_bchl}
\end{figure}

In case-2, BChl-based vegetation, as the major photosynthetic pigment, covers the surface of a planet. The BChls are assumed to emit the same degree of fluorescence intensity as the Earth's vegetation.
As shown in Figure~\ref{fig:abs}, fluorescence from BChls occurs in the wavelength range from 1000 to 1100\,nm.
In contrast to case-1, fluorescence emissions with 5 and $10F_\mathrm{fluor.}$ show strong features around 1050\,nm in almost all conditions in Figures~\ref{fig:ref_bchl} and \ref{fig:ref_bchl_2}.
Identifying the fluorescence on the Earth's vegetation level ($\lesssim F_\mathrm{fluor.}$) is still challenging even in the optimistic case, that is, (a) veg-only 0B.
The reflectivity between 1000 and 1050\,nm becomes slightly higher for mod-earth models with less surface vegetation coverage.
As shown in Figure~\ref{fig:diff_bchl}, the BChl organisms efficiently absorb photons and emit fluorescence with less absorption and scattering in the planetary atmosphere.
The fluorescence emissions from BChls that we assumed are invulnerable to blending with the steep increase in the reflectance by the VRE.
As a result, we found a more significant fluorescence contribution to the reflected light in case-2. 

Atmospheric properties, such as chemical compositions and cloud coverage, change
the fluorescence profile.
The water absorption is weak for wavelengths from 1000 to 1100\,nm.
If the major absorption bands of a photosynthetic pigment lie in wavelengths longer or shorter than 1000--1100\,nm, the presence of water vapor in the atmosphere complicates the detection of fluorescence emissions.
A strong absorption due to CH$_4$ in an oxygen-poor atmosphere also hides fluorescence near 1000\,nm (see the GJ667C models in Figure~\ref{fig:ref_bchl_2}).
The BChl organisms bearing BChl $b$ and their Stokes shift are ideal for detecting fluorescence in wavelengths longer than the characteristic wavelength of fluorescence from Chls.
Thus, fluorescence in the wavelength range of 1000 -1100\,nm could be a suitable biosignature for photosynthetic organisms, such as bacteriochlorophylls, on planetary surfaces unless they coexist with strong absorbers near 1000\,nm.

The VRE with a sharp rise in the reflectance is observed in the wavelength range from 1050 to 1100\,nm in case-2, as shown in Figure~\ref{fig:exoveg}.
Reflectance excess due to BChl fluorescence is 0.01--0.05 for the Modern Earth atmosphere models (see Figure \ref{fig:diff_bchl}), whereas that due to the VRE is 0.4--0.5 (0.1--0.15) for veg-only models (veg-land and mod-earth models). 
Bacteriochrolophylls' fluorescence causes a slight increase in reflectance around 1000 -1100\,nm compared to the VRE.
Such nonprominent fluorescence emission with a Gaussian shape in the wavelength different from the VRE feature can be extracted from the reflectance profile using data processing such as principal component analysis (PCA).
Photosynthetic organisms different from those around the Sun are expected to exhibit VRE and fluorescence features in different wavelengths.
Thus, not only spectral features due to atmospheric molecules but also the simultaneous detection of the VRE and the fluorescence will help identify traces of photosynthesis on an exoplanet.
Probably, when we found a possible signal of VRE, the fluorescence would be useful for further validation, because the VRE signal is stronger than the fluorescence one.

%%%%%%%%%%%%%%%%%%%%%%%%%%%%%%%%%%%%%%%%%%
%%%%%%%%%%%%%%%%%%%%%%%%%%%%%%%%%%%%%%%%%%

%%%%%%%%%%%%%%%%%%%%%%%%%%%%%%%%%%%%%%%%%%
%%%%%%%%%%%%%%%%%%%%%%%%%%%%%%%%%%%%%%%%%%

%%%%%%%%%%%%%%%%%%%%%%%%%%%%%%%%%%%%%%%%%%
%%%%%%%%%%%%%%%%%%%%%%%%%%%%%%%%%%%%%%%%%%

\subsection{Case-3: Anoxic World (without VRE)}

\begin{figure}
 \begin{center}
 \includegraphics[width=90mm]{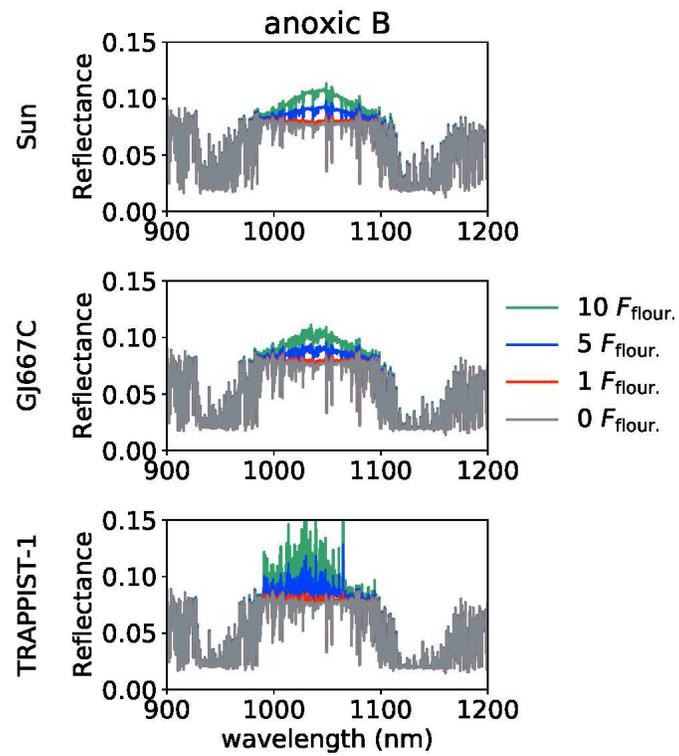}
 \end{center}
% \vspace*{65.0mm}
\caption{The reflectance of an Earth-like planet with an anoxic atmosphere and no land vegetation (anoxic B).}
 \label{fig:39}
\vspace*{300mm}
\end{figure}

In case-3, an Earth-like planet has the same reduced atmosphere as the Archean Earth at 3.9\,Ga. Anoxic bacteria with photosynthetic pigments such as bacteriochlorophylls may spread over the surface of a planet with a CO$_2$-rich atmosphere. Anoxic bacteria are assumed to live in the ocean and coast (i.e., $c_v = 0.72)$ and emit only fluorescence whose intensity is comparable to the standard emission from land plants, without the distinct reflectance of a vegetation surface.
Fluorescence emissions from anoxic bacteria adopt those from bacteriochlorophylls on the Earth. Figure\,\ref{fig:39} shows the reflectance of an Archean-Earth-like planet with BChl-based bacteria. In the reflection spectra, a strong water absorption appears around 950 and 1150\,nm. The relatively high reflectance across the wavelength range is mainly from the light reflected by the land.
We observe fluorescence emissions in the wavelength range between 1000 and 1100\,nm owing to the lack of light reflected from BChl-bearing oceanic bacteria, including the VRE feature.
Intense absorption in the stellar atmosphere enhances the apparent reflectance of a planet around TRAPPIST-1 (see also Figures~\ref{fig:ref_chl} and \ref{fig:ref_chl_2}, and Section~\ref{desc:appa}).

%%%%%%%%%%%%%%%%%%%%%%%%%%%%%%%%%%%%%%%%%%
%%%%%%%%%%%%%%%%%%%%%%%%%%%%%%%%%%%%%%%%%%

\section{Discussion} 
\label{desc:dis}

This study demonstrated reflectance with photosynthetic fluorescence on an Earth-like planet around the Sun and two M dwarfs.
This section reviews the biological processes of photosynthesis and then considers the future detection of biofluorescence on an exoplanet.
In Section~\ref{desc:physco}, we discuss the possible physiological conditions that enhance the fluorescence emissions on a planet based on our understanding of Chl fluorescence.
In Section~\ref{desc:false}, we discuss the possible false positive or negative detection of fluorescence (Section~\ref{desc:fp_general}), and the potential usage of the nonlinear photoresponse in fluorescence yield to excitation light intensity to distinguish between biofluorescence and the false positive/negative signals of fluorescence (Section~\ref{desc:nonl}).
Finally, in Section~\ref{desc:photo}, we show the fluorescence detection with telescopes.
We present the detectability of fluorescence from an Earth twin around a Sun-like star u sing the noise model for a LUVOIR-A-like mission (Section~\ref{desc:luvoir}), and
the remarkable enhancement in the reflectance due to the absorption lines of stars, which could be a promising feature for detection by high-dispersion spectroscopy, especially around ultracool stars (Section~\ref{desc:appa}).

\subsection{Possible Physiological Conditions for Supporting Fluorescence Detection}
\label{desc:physco}

This study adopted the typical fluorescence spectrum of Chl-containing plants and LH1--RC purified from BChl\,$b$-bearing purple bacteria.
The fluorescence spectrum of the LH1--RC complex suspended in buffer solution was measured under laboratory conditions with a low concentration of LH1--RC in the solution to avoid the reabsorption of fluorescence.
Cells having LH1-RC in vivo would result in an $\sim50$\,nm shift in the spectral peak wavelength toward longer wavelengths under dense conditions, because the reabsorption of fluorescence reduces the shorter-wavelength part of fluorescence.
A red-shifted fluorescence spectrum should still be observable because it is located within the atmospheric window. For the fluorescence intensity of vascular plants on the ground, we referred to the standard value ($F_{\rm fluor.}$) for the fluorescence model on exoplanets in our simulations. The possible detection of fluorescence emissions on exoplanets would require $\gtrsim 5 F_{\rm fluor.}$ with BChl (see Figures~\ref{fig:ref_bchl}, \ref{fig:ref_bchl_2}, and \ref{fig:39}). There are four potential factors that increase the fluorescence yield in photosynthetic organisms from the biophysical viewpoint of photosynthetic studies on existing phototrophs on the Earth:

\begin{enumerate}
%%%%
  \item Increasing Chl/BChl concentration per land area \par
A high concentration of Chls and BChls enhances their fluorescence intensity.
In general, the Chl/BChl concentration in a cell increases for capturing as many photons as possible under low light conditions. Fluorescence increases linearly with Chl/BChl concentration when cell density is low. In contrast, the fluorescence intensity reaches a saturation level in highly dense environments due to the reabsorption of fluorescence by cells~\citep{du2017response}.

%%%%
  \item Small spectral overlap between absorption and fluorescence\par
The large separation between the main absorption band and its fluorescence band increases the fluorescence intensity of concentrated cells.
In photosynthetic organisms, the excitation energy is transferred between Chls, and
the Chl fluorescence tends to be emitted from long-wavelength Chls (LWC), which has the reddest absorption band in a photosystem because the excess excitation energy is easily trapped at the lowest energy level. 
A redshift in the peak wavelength of fluorescence and a blueshift in absorption, which can be caused by the modification of the vibronic interactions of pigments between surrounding proteins and solvent, reduce the spectral overlap between fluorescence and absorption. 
The fluorescence emission from LWCs is red-shifted to over 50 nm from that of bulk Chls in some conditions. Although most plants have a small amount of LWCs in PSII and the Chl fluorescence is absorbed well under high Chl concentrations, far-red absorbable LWC contributing to PSII has been reported in some eukaryote algae~\citep{fujita2004710,wilhelm2006uphill,kotabova2014novel,wolf2018characterization,kosugi2020red}. These algae show a significant fluorescence emission at far-red-light wavelengths (700--800 nm) at room temperature, and some of them decrease the overlap% with absorption spectra
~\citep{fujita2004710,kosugi2020red}. 

%%%
  \item Low photosynthetic efficiency\par
%A high-efficiency fluorescence yield of cells enhances fluorescence emissions per site.
Photon loss in photosynthetic processes reduces the photon yield of fluorescence.
Excitation yield in PSII has increased throughout the evolutionary processes of photosystems.
 For example, the increase in light use efficiency in oxygenic photosynthesis on Earth was achieved by changing the light-harvesting antenna protein from the membrane superficial phycobilisome in cyanobacteria to the light-harvesting Chl binding protein in eukaryotic algae. 
Furthermore, the subsequent modification of LHCs achieved a higher photosynthetic quantum yield in the evolution process. The maximum excitation yield in PSII of vascular plants is estimated to be $\sim$ 0.9, whereas that of green algae and cyanobacteria is $\sim$ 0.8 and $\sim$ 0.6, respectively~\citep{schuurmans2015comparison}. Suppose phototrophs on an exoplanet are in the early stage of evolution. In that case, the expected fluorescence yield may be high to compensate for the low efficiency of photon yields in primitive photosynthesis.
%%%
  \item Suppression of heat dissipation\par
Photon loss by the heat dissipation in photosynthetic pigments suppresses the photon yield of fluorescence.
Heat dissipation occurs in the vibrational relaxation of excited pigment molecules, Chls, or accessory pigments such as carotenoids.
Additionally, light-dependent protection mechanisms to dissipate the excess light energy as heat are inherent in all the cyanobacteria, algae, and plants. The efficiency of heat dissipation largely depends on the molecular configuration and the environment of pigments binding to proteins. The energy conversion rate from light to heat in photosystems is crucial in estimating photosynthetic fluorescence on other planets.

\end{enumerate}
Therefore, the fluorescence yield in photosynthetic pigments should fluctuate over time due to photosynthetic activity and heat dissipation.

\subsection{Further Identification for Confirming Photosynthetic Fluorescence}
%\subsection{Possibility of False Positive/Negative Detection of Biological Fluorescence}
\label{desc:false}

\subsubsection{Potential false positive/negative of biological fluorescence detection from exoplanets}
\label{desc:fp_general}
%Same to the discussion on the possible extension of VRE to exoplanets, 
Photosynthetic pigments on an exoplanet may be different from those on Earth, and the wavelength relevant to fluorescence emission from exovegetation remains to be unknown.
%Therefore, potential false positive and negative detection of photosynthetic fluorescence should always be paid attention to sufficiently.
A possible fluorescence signal on other planets can be a false positive or negative detection of biological activities. 
Potential main sources causing false positive/negative could be surface reflectance or fluorescence from minerals on exoplanets.
Both Chl and BChl fluorescence in our study can be contaminated by mineral fluorescence, but it is not plausible to expect the fluorescent minerals to cover a fraction of a planetary surface comparable to Earth's vegetation as far as our knowledge of the Earth's environment.
Recently, solar-induced mineral luminescence (SML) has been extracted from SIF data obtained by remote sensing of the Earth~\citep{Kohler2021-jp}.
They revealed that about 10$\%$ of non-vegetated areas are weakly luminescent and speculated that luminescence came from some spots covered by carbonate with Mn$^{2+}$ and was comparable to SIF (or Chl fluorescence). However, those areas are negligible on the planetary scale.
%\cite{Smith2014-gi} 
%Thus, the Chl fluorescence can be affected by these emissions in terms of the SIF detection on the Earth.
On the other hand, mineral fluorescence could pollute, to an extent, fluorescence in near-infrared, which includes the BChl fluorescence.
For instance, silicate (e.g., pyroxene and olivine) shows a prominent absorption around 1000 nm caused by Fe$^{2+}$~\citep{bishop2019remote,Klima2011-jx,sunshine1998determining}. Its fluorescence could appear in a slightly longer wavelength from the absorption, whose energy corresponds to the Stokes shift, like other near-infrared fluorescent materials~\citep{Jackson2021-pn,Selvaggio2020-ak}.
While there are a variety of fluorescent minerals (e.g., fluorite, calcite, corundum),
we do not deny the possibility that the unexpectedly strong mineral fluorescence could be observed on exotic planets such as a carbide exoplanet~\citep{Allen-Sutter2020-jm} whose surface could be covered by diamond with lattice defects, e.g., due to nitrogen-vacancy center~\citep{Schirhagl2014-kh}.
To understand potential fluorescence features from surface components of an exoplanet, e.g., rocks and minerals, characterizing atmospheric features is helpful. Besides,
%Characterizing atmospheric features helps understand potential fluorescence features from surface components of an exoplanet, e.g., rocks and minerals. 
as mentioned so far, the simultaneous detection of vegetation reflectance (VRE) and fluorescence features could help identify photosynthesis.

%   Although there are a variety of fluorescent minerals (i.e., fluorite, calcite, corundum),
% We do not deny the possibility the strong mineral fluorescence could be observed on exotic planets such as a carbide exoplanet (DOI 10.3847/PSJ/abaa3e) which surface could be covered by diamond with nitrogen-vacancy center (10.1146/annurev-physchem-040513-103659)

\subsubsection{Nonlinear photoresponse in photosynthesis}
\label{desc:nonl}
Photosynthetic organisms regulate metabolic processes to maximize the use of available photons under light conditions and emit biological fluorescence by converting light energy via photochemical reactions.
The nonlinear response of the fluorescence yield to the excitation light intensity would be a clue to finding the presence of photosynthetic organisms. 
If a planet is in an elliptical orbit, the incident flux received by the planet from its host star varies with time.
Fluorescence emissions from nonbiological processes increase with incident light intensity. In contrast, a saturation level of the fluorescence intensity from biological activities, such as photosynthesis, exists
because the quantum yields of Chl fluorescence vary according to the light environment and atmospheric CO$_2$ concentrations. The quantum yields of Chl fluorescence are primarily involved in the reduction states of electron acceptors of photosystems for electron transports and excitation energy quenching by photoprotection mechanisms \citep[see][]{genty1989relationship,krause1991chlorophyll,baker2008chlorophyll}.
A sudden intense light can induce the reduction in the electron acceptors of PSII, where oxidation of water to generate O$_2$ occurs as a primary step in photosynthesis.
The presence of photoprotection mechanisms also modulates the quantum yields of Chl fluorescence. When dark- or dim-light-adapted leaves are suddenly irradiated with intense light, Chl fluorescence quantum yields rapidly increase by up to five times.
Accordingly, the relationship between fluorescence yield and excitation light intensity (i.e., the number of absorbed photons) provides a hint to explore the origin of fluorescence on a planet.

\begin{comment}
shoumetsu.murigaattaka.
\subsubsection{Inverse Correlation between Fluorescence and CO$_2$ Emissions}
\label{desc:co2}
The seasonal changes in the fluorescence and CO$_2$ emissions provide clues to identifying the biological fluorescence from a planet. 
Remote sensing on the Earth shows that the SIF is inversely related to the surface CO$_2$ flux because SIF shows active photosynthesis and CO$_2$ is drawn down during active photosynthesis.
The SIF shows peak values in a year over the northern hemisphere, and the time derivative of atmospheric CO$_2$ concentration becomes negative due to photosynthetic activity.
In winter, 
photosynthetic activity over the northern hemisphere is inhibited (i.e., the minimum SIF value is attained), and more CO$_2$ is released by respiration. If an inverse correlation exists between the seasonal changes in the time series of fluorescence and CO$_2$ emissions, it may be a biosignature. Note that high-spectral-resolution observations should separate the contribution of each hemisphere without averaging the whole sphere of a planet. 
\end{comment}

%%%%%%%%%%%%%%%%%%%%%%%%% Mock observation (begin) %%%%%%%%%%%%%%%%%%%%%%%%%%%%%%

\subsection{Detectability of Biological Fluorescence by Future Telescopes}
\label{desc:photo}

%In this section, we discuss the possible detection by future space-based and ground-based telescopes.
%In 4.3.1, the detectability of the fluorescence is demonstrated from an Earth's twin around a Sun-like star using the noise model for a LUVOIR-A-like mission.
%And then, in 4.3.2, the apparent enhancement in the reflectance around TRAPPIST-1, as seen so far, is shown further, which could be a hopeful clue for the detection by ground-based telescopes. 

\subsubsection{The Earth-Sun System as an Earth Twin in a LUVOIR-A-Like Mission}
\label{desc:luvoir}

\begin{figure}
 \begin{center}
 \includegraphics[width=140mm]{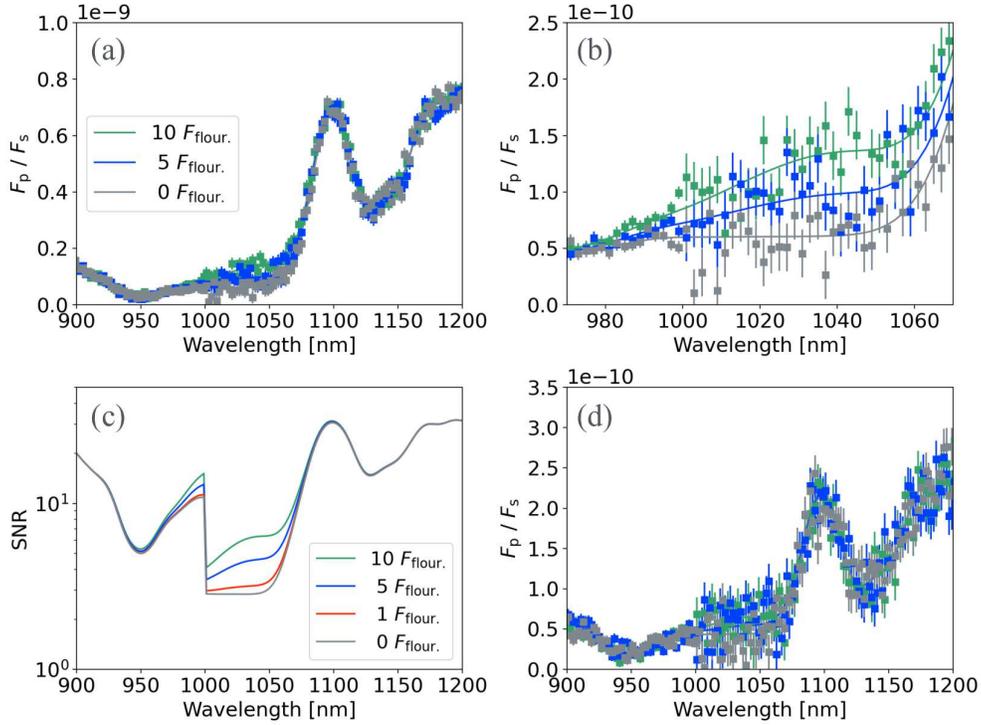}
 \end{center}
\caption{Simulated spectrum with the biological fluorescence on a second Earth around a Sun-like star at 10 pc from the Earth, assuming a LUVOIR-A-like space telescope. (a--c) The results from the veg-only 0B model and (d) $F_{\rm p} / F_{\rm s}$ with the veg-land 0B model. (a) $F_{\rm p} / F_{\rm s}$ with 9000 hours of observation time. The solid line shows $F_{\rm p} / F_{\rm s}$ and the error bar indicates the noise at each wavelength. (b) A magnification of $F_{\rm p} / F_{\rm s}$ in (a).
(c) The SNR in (a). }
 \label{fig:obs_bchl1}
\end{figure}

\begin{figure}
 \begin{center}
 \includegraphics[width=140mm]{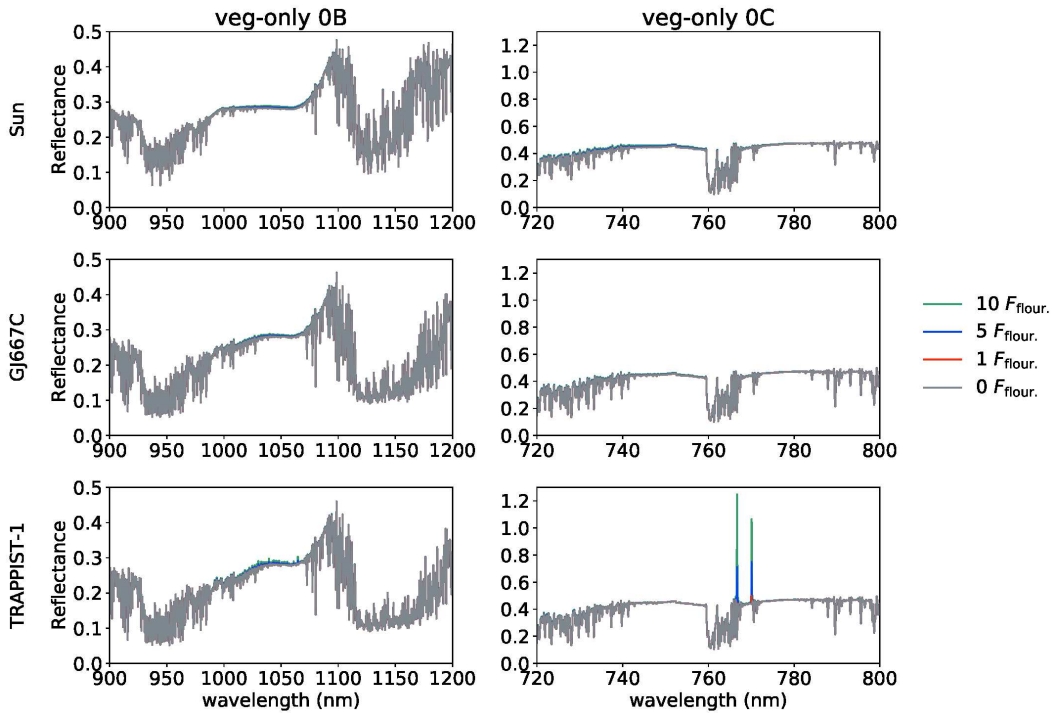}
 \end{center}
\caption{The effect of cloud on the reflectance with veg-only 0B and 0C models. The models are the same as the veg-only 0B in Figure~\ref{fig:ref_bchl} and the veg-only 0C in Figure~\ref{fig:ref_chl} but with cloud coverage.}
 \label{fig:ref_bchl_cloud}
%\vspace*{-3mm}
\end{figure}

We investigated the detectability of fluorescence from an Earth twin around a Sun-like star at 10\,pc from the Earth, assuming a LUVOIR-A-like space telescope.
Figure~\ref{fig:obs_bchl1} presents the simulated spectra of a second Earth around a Sun-like star at 10\,pc with the biological fluorescence.
We applied the noise model used in \cite{robinson2016characterizing} and \cite{kopparapu2021nitrogen}, which accounts for planet photons, stellar photon noise, and background noise, e.g., zodi, exozodi, read-out, and dark current noises with the throughput assuming the LUVOIR-A telescope.
The parameters and the formalism used in this paper are presented in Appendix B. 
Figures~\ref{fig:obs_bchl1}(a--c) show the results of the most optimistic model for the fluorescence signal (veg-only 0B) from the Earth-Sun system observed from 10 pc with a 15 m space telescope. The original data are the same as those of the Sun in Figure~\ref{fig:ref_bchl}(a).
In Figure~\ref{fig:obs_bchl1}(a), $F_{\rm p} / F_{\rm s}$ observed at the telescope for each wavelength bin is shown as solid lines, with the random noise as the 1$\sigma$ error bars for each bin, in 9000 hours of exposure time, where $F_{\rm p}$ is the reflected light from the planet and $F_{\rm s}$ is the starlight.
Figure~\ref{fig:obs_bchl1}(b) depicts a magnification of the spectrum in Figure~\ref{fig:obs_bchl1}(a).
Some error bars are outside the solid line, 
but the spectral feature of fluorescence emission is recognizable for each case in the figure.
Figure~\ref{fig:obs_bchl1}(c) shows the SNR with the same observation time as that in Figure~\ref{fig:obs_bchl1}(a).
The difference between 0 and 5 $F_{\rm {fluor.}}$ is larger than 1$\sigma$. % may be detectable with medium-resolution direct imaging spectroscopy.
To detect the fluorescence with 3$\sigma$ error, $\sim$ 50000 hours of exposure time are required, and with 5$\sigma$, $\sim$ 100000 hours, ten years, are expected (not shown in figures). 
Thus, fluorescence detection would require years for observation, even by the LUVOIR-A-like space telescope, and it is extremely challenging to observe one target.
In less optimistic models, namely, the veg-land 0B model around the Sun in Figure~\ref{fig:ref_bchl}(b), the detection of fluorescence signals is even more challenging, as shown in Figure~\ref{fig:obs_bchl1}(d).
As discussed in Section 3.2, the fluorescence in mod-earth 0B is difficult to identify. Moreover, cloud coverage obscures the VRE features as well as atmospheric features on exoplanets \citep{seager2005vegetation,tinetti2006detectability,kaltenegger2007spectral}.
The reflectance in Figure~\ref{fig:ref_bchl_cloud} indicates how clouds suppress the fluorescence signal.
Even in the most optimistic model,
the fluorescence in the reflectance is significantly reduced and can hardly be observed.
In the mod-earth model, 
it is impossible to identify the fluorescence signals.
The only possible way to observe surface vegetation with significant cloud coverage, except for atmospheric gases, would be the VRE ($\sim$0.1 in reflectance in the optimistic model).
Thus, the existence of water clouds that are expected in Earth-like planets with surface water seems to be critical for fluorescence detection.
However, around TRAPPIST-1, as the relevant argument was shown in Session~\ref{desc:appa}, we found that the Chl fluorescence in the K $\rm{I}$ lines was insensitive to the coverage by Earth clouds, which could be an advantage in the Chl detection over BChl one.

%The small planet-star separation of a planet in the habitable zone around M dwarfs could be inside of the inner working angle (IWA) with direct imaging and difficult to be resolved even by future space telescopes.
%Therefore, the detectability of fluorescence from an Earth's twin around a Sun-like star at 10 pc from the future space telescope, rather than around GJ667C and TRAPPIST-1, is presented in Figure~\ref{fig:obs_bchl1}. 

The fluorescence feature would be poorly determined with 900 hours of exposure time with 1$\sigma$ errors, whereas the VRE feature can be identified.
Even for a LUVOIR-A-like space telescope, an enormous observational time would be needed to identify the fluorescence in addition to the VRE with more confidence for detecting traces of photosynthesis. 
We also investigated the detectability of fluorescence by a space telescope with a different diameter. 
A 6 m space telescope is recommended for future space missions, according to Astro2020 Decadal Survey.
With a 6-m diameter, $\sim$ 300,000 hours of observation time are required to identify fluorescence.
When we adopt a 30 m space telescope with 1$\sigma$ errors, the required exposure time is reduced to $\sim$ 800 hours.
Furthermore, one of the background noises, i.e., the readout noise, can be suppressed with data processing because of increasing reads in an exposure as implemented for H2RG infrared detectors~\citep[e.g.,][]{Brandt_2018_CHARIS,kuzuhara2018performance}.
When the readout noise is assumed to be zero all over the wavelengths, the required observation times are reduced to $\sim$ 250,000, $\sim$ 7,000 and $\sim$ 500 hours with the 6-, 15-, and 30-m diameters.

%%%%%%%%%%%%%%%%%%%%%%%%%%%%%%%%%%%%%%%%%%
%%%%%%%%%%%%%%%%%%%%%%%%%%%%%%%%%%%%%%%%%%

%%%%%%%%%%%%%%%%%%%%%%%%%%%%%%%%%%%%%%%%%%
%%%%%%%%%%%%%%%%%%%%%%%%%%%%%%%%%%%%%%%%%%

%%%%%%%%%%%%%%%%%%%%%%%%%%%%%%%%%%%%%%%%%%
\begin{figure}
 \begin{center}
 \includegraphics[width=180mm]{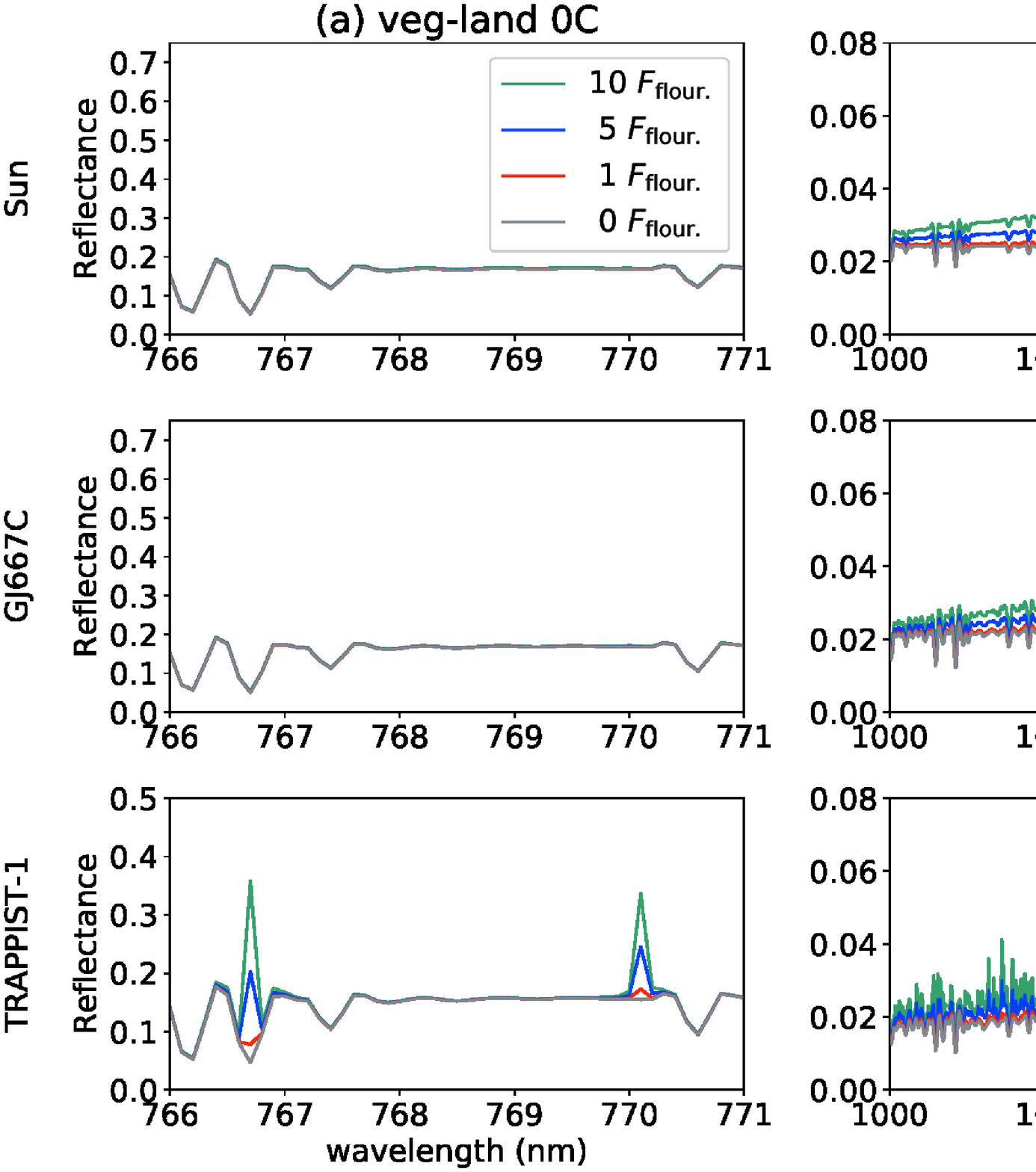}%{reflect_zoom_trap.eps}
 \end{center}
 \vspace*{2.0mm}
\caption{The apparent enhancement of fluorescence in reflectance due to stellar absorption around the three template stars: (a) veg-only 0C model (Figure~\ref{fig:ref_chl}), (b) veg-only 0B model (Figure~\ref{fig:ref_bchl}), and (c) anoxic B model (Figure~\ref{fig:39}).}
 \label{fig:trap}
\vspace*{-3mm}
\end{figure}
%%%%%%%%%%%%%%%%%%%%%%%%%%%%%%%%%%%%%%%%%%

\subsubsection{Apparent Enhancement in Fluorescence around Ultracool Stars and Possible Detection with High-Dispersion Spectroscopy}
\label{desc:appa}

Figure~\ref{fig:trap} shows the contribution of fluorescence around three host stars.
%Around TRAPPIST-1, the apparent enhancement in reflectance induced by fluorescence is significant compared to around the other two stars because TRAPPIST-1 shows strong absorption features.
%The reflectance increases by the fluorescence emission and the low incoming flux due to stellar absorption. 
Around TRAPPIST-1 the apparent enhancement in reflectance induced by fluorescence is significant compared to around the other two stars because TRAPPIST-1 has strong absorption features spanning the wavelengths of the fluorescence.  Within the TRAPPIST-1 stellar absorption features, reflected light from the planet is reduced, allowing the fluorescence emission to become a much larger fraction of the outgoing flux (reflected + fluorescence) at these wavelengths. This is analogous to the methodology of SIF detection with remote sensing observations and the retrieval processes by determining how much the fluorescence influences the Fraunhofer lines~\citep{maier2004sun}.
These spectroscopic features may be widely used for fluorescence detection around ultracool stars.

Figure~\ref{fig:trap}(a) shows that the reflectance is highly enhanced due to the absorption lines of K $\rm{I}$ in the stellar spectrum of TRAPPIST-1, which is not affected by water clouds (Figure~\ref{fig:ref_bchl_cloud}).
The degree of enhancement for each line depends on the atmospheric compositions of an Earth-like planet. 
Figure~\ref{fig:trap}(b,c) presents a spiky feature due to absorption of FeH and VO, as commonly observed around ultracool stars.
Therefore, observing the possible fluorescence signal with high spectral resolution using extremely large ground telescopes would be worthwhile.

\begin{comment}
{\com [

https://iopscience.iop.org/article/10.1086/499583/pdf

https://iopscience.iop.org/article/10.3847/1538-4357/aa7fea/pdf
]}

\end{comment}

\section{Conclusions} 

In this paper, we explored fluorescence from photosynthesis as a biosignature on an exoplanet for future observations in great detail and identified the situations in which the signal could be enhanced, and the regions of the spectrum where fluorescence from chlorophylls and bacteriochlorophylls could be most detectable for Earth-like planets around different stars. We also described how we could enhance the possibility to more definitively detect the action of photosynthesis. 
For direct imaging observations, however, we found that the detection of fluorescence emissions would be extremely challenging to observe and especially not feasible for the planned 6m space telescope. More details are provided as follows.

%This study examined the detectability of biofluorescence from photosynthesis on an exoplanet for future observations. %Please note that the simple past tense is used to describe your actions in a study, not the present perfect.
We considered fluorescence emissions from Chl- and BChl-based vegetation in a clear-sky condition on an Earth-like planet around the Sun and two M dwarfs (GJ667 C and TRAPPIST-1). 
% general
Chl- and BChl-based leaves show a VRE in wavelengths around 700--750 and 1000--1100\,nm.
The fluorescence emissions from Chls and BChls occur at wavelengths from 650 to 800\,nm and 1000 to 1100\,nm, corresponding to the longest Q absorption band of each pigment. The two peaks of Chl fluorescence at 680 and 740\,nm arise from the PSII and PSI, respectively.
Thus, atmospheric absorption bands, such as H$_2$O, CH$_4$, O$_2$, and O$_3$, and the VRE could be overlapped with the fluorescence emissions from Chls and BChls.
 Chl fluorescence emission from PSI is blended with the steep VRE feature.
Fluorescence emitted from PSII on an Earth-like planet is the most promising feature for observation, 
but it may also be reduced by nonphotochemical quenching processes and reabsorption of photons by surrounding Chls.
Conversely, the fluorescence emitted from BChls is not suppressed by the sharp increase in the reflectance due to the VRE and atmospheric absorption by, for example, water vapor, except for CH$_4$ absorption around 1000\,nm.
Therefore, the BChl fluorescence in the wavelength range of 1000--1100\,nm, rather than Chl fluorescence, may be a more promising biosignature from photosynthetic organisms on a planetary surface.
In both cases of Chl- and BChl-based vegetation, the simultaneous detection of the VRE and fluorescence is significant for identifying photosynthetic activity on an exoplanet, because we do not know exactly what kind of vegetation exists in the planet in principal and we need more information for further validation to identify the trace of photosynthesis.
If BChl-bearing photosynthetic bacteria inhabit water without any leaf or tree structures, the fluorescence spectrum is the only surface reflectance feature that can be used to access such underwater photosynthetic organisms, although the fluorescence signal would be reduced according to the opacity of overlying liquid water.

% 4.1
Based on our understanding of photosynthesis,
the intensity of fluorescence is lower in photosynthetic bacteria compared to land plants. 
Here, we presented four factors that enhance the fluorescence emission for possible detection of biological fluorescence on an exoplanet: (1) increase in Chl/BChl concentration per land area, (2) small overlap of absorption and fluorescence spectrum, (3) low photosynthetic efficiency, and (4) suppression of heat dissipation.
% 4.2
This study assumed a linear photoresponse of fluorescence to excitation light intensity.
If a planet is on a large elliptical orbit and the telescope has sufficient sensitivity to temporally resolve changes in fluorescence as a function of time, the nonlinear photoresponse from the biological fluorescence can be identified.
%Moreover, the inverse relationship between the CO$_2$ concentration and biological fluorescence can confirm the biosignature on an exoplanet. 
% 4.3
Assuming a LUVOIR-A-like mission, an enormous duration (around 9000 hours) would be required to detect the BChl fluorescence emission, whose fluorescence yield is 5--10 times larger than that of vegetation on Earth in the optimistic cases for an Earth-Sun twin at a distance of 10 pc from the Earth. 
In addition, the cloud coverage significantly affects the detection of fluorescence as well as other spectral features because the cloud more strongly obscures fluorescence emissions than the VRE feature. 
Interestingly, the fluorescence in the reflectance was found to be remarkably enhanced in all three cases around TRAPPIST-1 because of its strong absorption in the stellar atmosphere, like the SIF detection by remote sensing using Fraunhofer lines. 
The reflectance excess due to K $\rm{I}$ absorption and VO/FeH absorption can be a promising feature for characterizing the fluorescence around ultracool stars in Chl and BChl cases. 
Note that Chl fluorescence in K $\rm{I}$ lines was still prominent with water clouds.

%Thus, follow-up observations using large ground-based telescopes, with the high-spectral resolution, would be effective for further identification.

% matome

Thus, one of the most important future works would be the mock observation assuming a 30 m class ground-based telescope to investigate how the apparent enhancement in reflectance due to stellar absorption could help the fluorescence detection around ultracool stars.
In addition, to better support the future detection of fluorescence emissions on an exoplanet, further studies are required from various perspectives.
For example, planetary spectra for a wide range of atmospheric and surface conditions consistent with biological fluorescence emission should be estimated and tested using radiation transfer calculations because our studies considered still-limited conditions.
Moreover, we need to conduct simulations on how the fluorescence is observed on an exoplanet when a global SIF map data from remote sensing of the Earth are applied.
Also, experimental validation of prominent NIR fluorescence emissions is needed in some species of photosynthetic organisms and conditions.

\acknowledgments
We would like to thank one anonymous reviewer for constructive comments to improve the paper.
We also thank Tatsuya Miyauchi, Haruki Oshio, Yu Someya, Tomoki Kiyono, and Masanori Takeda for fruitful discussions at NIES on SIF detection by remote sensing, which led to the draft idea of this study, and Kouki Hikosaka (Tohoku University) and Hibiki Noda (NIES) for further discussions and for introducing SIF identification by remote sensing.
The data for the LUVOIR noise model was helpfully provided by Geronimo Villanueva and Ravi Kopparapu (NASA/Goddard).
Y.H. and N.N. were supported by a Grant-in-Aid for Scientific Research on Innovative Areas (JSPS KAKENHI grant number 18H05439).
PyAstronomy (\url{https://github.com/sczesla/PyAstronomy}) was used in mock observations assuming a space telescope.
In several cases, numerical data were extracted from figures in published papers using WebPlotDigitizer (\url{https://automeris.io/WebPlotDigitizer/}).

\appendix

\section{Empirical Rayleigh Scattering}

The effect of Rayleigh scattering is implemented empirically as follows~\citep{bucholtz1995rayleigh}:
\begin{eqnarray}
  \label{eqn:ray1}
  \tau_\mathrm{R} (\lambda) &=& \beta_\mathrm{s} (\lambda) \frac{T_\mathrm{s}}{P_\mathrm{s}} \int_{0}^{z'} \frac{P(z)}{T(z)} dz,
\end{eqnarray}
where $\tau_\mathrm{R}$ is the Rayleigh optical depth at altitude $z'$; $T (z)$ and $P (z)$ are the temperature and pressure at $z$, respectively.
We adopted the $T-P$ profile in the U.S. standard atmosphere 1976 from 0 to 60\,km to compute the Rayleigh scattering cross-section in the atmosphere of an Earth-like planet. The actual $T-P$ profile in the atmosphere of an Earth-like planet around a star other than the Sun is quite different from that in the Earth's atmosphere. Rayleigh scattering, however, has a negligible effect on the transmittance at wavelengths from 600 to 1100\,nm ($\approx$ 6 $\%$ in transmittance at 600 nm, reducing with increasing wavelength, and then $<$ 1 $\%$ at 1100\,nm for an Earth-like planet around the Sun, for instance), which is closely related to the fluorescence from Chls and BChls.
$T_\mathrm{s}$ and $P_\mathrm{s}$ are the temperature and pressure at standard conditions on Earth, respectively ($T_\mathrm{s}$ = 288.15\,K and $P_\mathrm{s}$ = 1013.25\,mbars). The total Rayleigh volume-scattering coefficient $\beta_s$ is expressed as:
\begin{equation}
  \label{eqn:ray2}
  \beta_s (\lambda) = A\lambda^{ -B-C\lambda-D/\lambda}, 
\end{equation}
where the coefficients $A, B, C$, and $D$ are empirically determined (see Table 3 in \cite{bucholtz1995rayleigh}). 
%%%%%%%%%%%%%%%%% methods for mock observations %%%%%%%%%%%%%%%%

%%%%%%%%%%%%%%%%%%%%%%%%%%%%%%% table for all setting %%%%%%%%%%%%%%%%%%%%%%%%%%%%%%% 
\begin{table}
\centering
\scalebox{1.0}{
\begin{tabular}{llc} \hline
    Parameter & Description & Adopted Value \\ \hline \hline
      $D$  & Mirror Diameter & 6, 15, 30 m  \\ 
      $C$  & Raw Contrast & $10^{-10}$ \\ %~\citep{meadows2018habitability,kopparapu2021nitrogen} \\ 
      $R$  & Instrumental spectral resolution & 70\\ %~\citep{meadows2018habitability,robinson2016characterizing} \\
      $T_\mathrm{Tele}$  & Accounts for light lost due to contamination & 0.95 \\ %~\citep{kopparapu2021nitrogen}  \\
      & and inefficiencies in the main collecting area & \\ 
      $T_\mathrm{read}$  & Read-out efficiency & 0.75 \\ %~\citep{kopparapu2021nitrogen}  \\  
      $T_\mathrm{QE}$  & Raw quantum efficiency & 0.9 \\ %~\citep{kopparapu2021nitrogen}  \\  
      $f_\mathrm{pa}$  & Fraction of planetary light that falls within photometric aperture & 1 \\%~\citep{kopparapu2021nitrogen} \\  
      $X$  & Width of photometric aperture as multiple of $\lambda / D$ & 0.61 arcsec \\ %~\citep{kopparapu2021nitrogen} \\
      %$F_{0,V}$ & Standard zero-magnitude V-bandspecific flux & 3.63 $\times 10^{-8} Wm^{-2} \mu m^{-1}$ \\ %~\citep{robinson2016characterizing}\\
      %$M_{z,V}$ & V-band zodicacal light surface brightness & 23 mag arcsec$^{-2}$ \\ %~\citep{robinson2016characterizing} \\ 
      %$M_{ez,V}$ & V-band exozodicacal light surface brightness & 22 mag arcsec$^{-2}$\\ %~\citep{robinson2016characterizing} \\ 
      $N_\mathrm{ez}$ & Number of Exozodis & 4.5 \\ %~\citep{kopparapu2021nitrogen} \\
      $D_{e-}$ & Dark current (UVIS/NIR) & 3E-5/2E-3 $e^{-}/s$ \\ %~\citep{kopparapu2021nitrogen} \\
      $R_{e-}$ & Read noise per pixel (UVIS/NIR)\tablenotemark{$a$} & 0/2.5 $e^{-}$ \\ %~\citep{kopparapu2021nitrogen}\\
      %$\Delta N_{hpix}$ & number of horizontal/spatial pixels for dispersed spectrum & 3 \\ %~\citep{robinson2016characterizing} \\
      %$\Delta t_{max}$ & Detector maximum exposure time & 1 hour \\ %~\citep{robinson2016characterizing}\\
     $\theta_\mathrm{IWA}$ & Inner working angle of the coronagraph as multiple of $\lambda / D$  & 3 \\
     $\lambda_{0}$ & Diffraction limit at the wavelength & 500 nm 
      \\\hline
\end{tabular}
}
  \caption{Parameters for simulations based on a LUVOIR-A-like mission.}
  \tablenotetext{$a$}{Taken from the Planetary Spectrum Generator for \textit{LUVOIR}/A-VIS and A-NIR, which is maintained by NASA (\url{https://psg.gsfc.nasa.gov/instrument.php}). }
      \label{tab:settings}
\end{table}
%%%%%%%%%%%%%%%%%%%%%%%%%%%%%%%%%%%%%%%%%%%%%%%%%%%%%%%%%%%%%%%%%%%%%%%%%%%%%%%%%%%%%%%%%%%%%

\section{LUVOIR Noise Model}

We implemented a noise model assuming a LUVOIR-A-like mission.
The formalism and the parameters are based on~\cite{robinson2016characterizing}, 
but, as shown in Table 2, we updated some parameters (with several treatments) following \cite{kopparapu2021nitrogen} for our simulations with the LUVOIR-A telescope.

The total noise in the observation $C_\mathrm{total}$ is calculated by:
\begin{eqnarray}
  \label{eqn:LUVOIR_c_tot}
  C_\mathrm{total} &=& C_\mathrm{p} + C_\mathrm{s} + C_\mathrm{b}, 
\end{eqnarray}
where $C_\mathrm{p}$ is the number of planet photons, $C_\mathrm{s}$ is the stellar photon noise (leakage through the coronagraph), and $C_\mathrm{b}$ is the background noise, which is the sum of zodi $C_\mathrm{z}$, exozodi $C_\mathrm{ez}$, dark current $C_\mathrm{D}$, and readout noise $C_\mathrm{R}$. The internal thermal noise is ignored because the thermal contribution is negligible in our wavelengths of interest.
%All variables are functions of wavelength $\lambda$.
Note that the noise in Equation \ref{eqn:LUVOIR_c_tot} corresponds to variance rather than the standard deviation. 
The noise count is expressed as:
\begin{eqnarray}
  \label{eqn:LUVOIR_noise}
  C_\mathrm{noise} &=& \sqrt{C_\mathrm{p}+C_\mathrm{s}+2C_\mathrm{b}}
\end{eqnarray}
where the double $C_\mathrm{b}$ accounts for the on-off observation with and without the planet.
The on-off observation corresponds to the subtraction of point spread functions of a central star. 
$\mathrm{S}/\mathrm{N}$ for each wavelength $\lambda$ is defined by:
\begin{eqnarray}
  \label{eqn:LUVOIR_sn}
  S/N &=& \frac{C_\mathrm{p}}{C_\mathrm{noise}}.
\end{eqnarray}
The $F_\mathrm{p}$ and $F_\mathrm{s}$ are now defined to be the reflected light from a planet and the stellar flux acquired by the telescope at a wavelength (bin) $\lambda$. 
%When observing $F_\mathrm{p}/F_\mathrm{s}$, where $F_\mathrm{p}$ is the reflected light from a planet at a telescope at wavelength (bin) $\lambda$ and $F_\mathrm{s}$ is the stellar flux at the telescope,
When observing $F_\mathrm{p}/F_\mathrm{s}$, the $1\sigma$ error at $\lambda$ is given as:
\begin{eqnarray}
  \label{eqn:sigma}
  \sigma(\lambda) &=& \frac{F_\mathrm{p}}{F_\mathrm{s}} \frac{1}{\mathrm{S}/\mathrm{N}}.
\end{eqnarray}

The end-to-end throughput for planetary fluxes is calculated as:
\begin{eqnarray}
  \label{eqn:LUVOIR}
  T_\mathrm{total} &=& T_\mathrm{Tele} T_\mathrm{cor} T_\mathrm{opt} T_\mathrm{read} T_\mathrm{QE},
\end{eqnarray}
where $T_\mathrm{Tele}$ is an account for light lost due to contamination and inefficiencies in the main collecting area, $T_\mathrm{read}$ is the read-out efficiency, and $T_\mathrm{QE}$ is the raw quantum efficiency for the detector.
The coronagraphic $T_\mathrm{cor}$ and the optical $T_\mathrm{opt}$ throughputs are the same as in Figure 9 in \cite{kopparapu2021nitrogen}.

% Kuzuhara-san's zodis model is shown here.
We updated the formalism on noise from zodis, exozodis, and readout as follows;
In~\cite{robinson2016characterizing}, the spectral shape of zodis (exozodis) was assumed to be equal to that of the Sun (the host star). 
Instead, we explicitly adopt the normalized reflectance on solar zodis, $\tilde{R}_{\odot,\lambda}$, in the model to better account for the zodical light in a exoplanetary system.
We calculate $\tilde{R}_{\odot,\lambda}$ by tracing the spectral data from observations of the zodical light (see Figure 8 in~\cite{kawara2017ultraviolet} and Figure 10 in~\cite{tsumura2010observations}) with the normalization in the $V$ band.
Using $\tilde{R}_{\odot,\lambda}$, the noise from zodis is expressed as:
\begin{eqnarray}
  \label{eqn:luvoir_zodis}
  C_\mathrm{z} &=& \frac{\pi\lambda^2 D^2}{4hcR} \frac{F_{\odot,\lambda}(1 \mathrm{au})}{F_{\odot,V}(1 \mathrm{au})} \tilde{R}_{\odot,\lambda}F_{0,V}10^{-M_{\mathrm{z},V}/2.5} T_\mathrm{total} \Omega \Delta t_\mathrm{exp},
\end{eqnarray}
where $F_{\odot,\lambda}$ is the solar flux density at $\lambda$, $F_{\odot,V}$ is the solar flux density in the $V$ band, $h$ is the Planck constant, $c$ is the speed of light, $M_{\mathrm{z},V}$ = 23 mag arcsec$^{-2}$ is the $V$-band zodical-light surface brightness, and $\Delta t_\mathrm{exp}$ is the exposure time. 
The circular photometry aperture size is expressed as $\Omega=\pi(X\lambda/D)^2$.
%~\citep{meftah2018solar}.
Assuming the exozodis's reflectance to be the same as $\tilde{R}_{\odot,\lambda}$, the noise from exozodis is written as:
\begin{eqnarray}
  \label{eqn:luvoir_exozodis}
  C_\mathrm{ez} &=& \frac{\pi\lambda^2 D^2}{4hcR} \left(\frac{1 \mathrm{au}}{r} \right)^2 \frac{F_{\mathrm{s},\lambda}(1 \mathrm{au})}{F_{\mathrm{s},V}(1 \mathrm{au})} \frac{F_{\mathrm{s},V}(1 \mathrm{au})}{F_{\odot,V}(1 \mathrm{au})} \tilde{R}_{\odot,\lambda}F_{0,V}N_\mathrm{ez}10^{-M_{\mathrm{ez},V}/2.5} T_\mathrm{total} \Omega \Delta t_\mathrm{exp},
\end{eqnarray}
where $F_{\mathrm{s},\lambda}$ is the stellar flux density at $\lambda$, $F_{\mathrm{s},V}$ is the stellar flux density in the $V$ band, and $r$ is the distance between the planet and the parent star. $M_{\mathrm{ez},V}$ = 22 mag arcsec$^{-2}$ is the $V$-band exozodical light surface brightness.
Even if the original treatment of exozodical light is adopted, our results do not significantly vary. 
%in our paper are not significantly affected, 
%but the treatment has a negligible effect when the contribution from the (exo-)zodical light is significant.
%but our adopted treatment is not negligible when the (exo-)zodical light is significant. 
We calculate the read-out noise ($C_\mathrm{R}$) to be $C_\mathrm{R}=N_\mathrm{pix}N_\mathrm{read}R_\mathrm{e^-}^2$ instead of $C_\mathrm{R}=N_\mathrm{pix}N_\mathrm{read}R_\mathrm{e^-}$ in~\cite{robinson2016characterizing} to more realistically incorporate the noise propagation,
%for reproduction of a reasonable noise level,
where $N_\mathrm{pix}$ is the number of contribution pixels, $N_\mathrm{read}$ is the number of reads at each observation, and $R_\mathrm{e^-}$ is the read noise count.

%\section{Word counter (removed later)}
%This document has {\WordCount} words.

\bibliography{reference}{}
\bibliographystyle{aasjournal}

\end{document}